\documentclass[aps,superscriptaddress,amsmath,amssymb,floatfix,twocolumn,showpacs,amsfonts,longbibliography]{revtex4-1}
\usepackage{times}
\usepackage[varg]{txfonts}
\usepackage{textcomp}
\usepackage{graphicx}
\usepackage{subfigure}
\usepackage{tabu}
\usepackage{color}
\usepackage[colorlinks=true,citecolor=blue,urlcolor=blue,linkcolor=blue,hyperindex]{hyperref}
\usepackage{braket}
\usepackage{overpic}
\usepackage{amssymb}

\def\kk{\mathbf{k}}
\def\qq{\mathbf{q}}

\def\non{\nonumber}

\allowdisplaybreaks

\begin{document}

\title{Magnon-phonon coupling effects on the indirect K-edge resonant inelastic X-ray scattering spectrum of a 2D Heisenberg antiferromagnet}
\author{Zijian Xiong}
\affiliation{State Key Laboratory of Optoelectronic Materials and Technologies, School of Physics, Sun Yat-Sen University, Guangzhou 510275, China}

\author{Trinanjan Datta}
\email[Corresponding author:]{tdatta@augusta.edu}
\affiliation{Department of Chemistry and Physics, Augusta University, 1120 15$^{th}$ Street, Augusta, Georgia 30912, USA}
\affiliation{State Key Laboratory of Optoelectronic Materials and Technologies, School of Physics, Sun Yat-Sen University, Guangzhou 510275, China}

\author{Kenneth Stiwinter}
\affiliation{Department of Chemistry and Physics, Augusta University, 1120 15$^{th}$ Street, Augusta, Georgia 30912, USA}

\author{Dao-Xin Yao}
\email[Corresponding author:]{yaodaox@mail.sysu.edu.cn}
\affiliation{State Key Laboratory of Optoelectronic Materials and Technologies, School of Physics, Sun Yat-Sen University, Guangzhou 510275, China}

\date{\today}

\begin{abstract}
We compute the effects of magnon-phonon coupling on the indirect K-edge bimagnon resonant inelastic x-ray scattering (RIXS) intensity spectrum of a square lattice Heisenberg antiferromagnet. We analyze the effects of competing nearest and next--nearest magnetic and magnon-phonon coupling interaction in the RIXS spectrum, for both the antiferromagnetic (AF) and the collinear antiferromagnetic (CAF) phases of the model. Utilizing the Dyson-Maleev representation of spin operators, the Bethe-Salpeter ladder approximation scheme for the bimagnon interacting channel, and considering the lowest order magnon-phonon-magnon scattering interaction we highlight distinct features in the X-ray spectrum. Considering damping effects, arising due to the presence of phonons, we find that in the AF phase the RIXS intensity spectrum attains a maximum value primarily localized around the K $\left(\pm\frac{\pi}{2}, \pm \frac{\pi}{2}\right)$ - point. For the CAF phase the intensity is broadly distributed with a significant scattering intensity located around the Y $\left(\pm\frac{\pi}{2}, 0\right)$ - point. Furthermore, in the CAF phase for suitable anisotropy, nearest-, and next-nearest neighbor interaction parameters the phonon effects can manifest itself as a distinct peak both below and above the bimagnon peak. Such a feature is in contrast to the antiferromagnetic spectrum where the effect due to the phonon peak was located consistently beyond the bimagnon peak in the high energy end of the spectrum. Additionally, in the CAF phase we find the RIXS bimagnon-phonon spectrum to be more sensitive to anisotropy compared to its antiferromagnetic counterpart. We conclude that the ultimate effect of magnon-phonon effects in the indirect K-edge RIXS spectrum, in both the antiferromagnetic and the collinear antiferromagnetic phase, is an observable effect.
\begin{description}
\item[PACS number(s)] 78.70.Ck, 75.25.-j, 75.10.Jm
\end{description}
\end{abstract}

\maketitle
%%%%%%%%%%%%%%%%%%%%%%%%%%%%%%%%%%%%%%%%%%%%%%%%%%%%%%%%%%%%%%%%
\section{Introduction}
%%%%%%%%%%%%%%%%%%%%%%%%%%%%%%%%%%%%%%%%%%%%%%%%%%%%%%%%%%%%%%%%
Improved X-ray instrumentation resolution coupled with advanced X-ray synchrotron radiation sources have established resonant inelastic X-ray scattering (RIXS) as a significant experimental tool to explore condensed matter systems \cite{RevModPhys.83.705,DEAN20153}. Experimental ~\cite{PhysRevLett.100.097001,PhysRevLett.102.167401,PhysRevLett.103.047401,PhysRevLett.104.077002,PhysRevLett.105.157006,PhysRevLett.107.107402,
NatPhys.7.725,PhysRevLett.108.177003,PhysRevB.85.214527,NatMater.11.850,NatMater.12.1019,PhysRevLett.103.107205,Nature.485.82,PhysRevLett.110.265502,PhysRevLett.110.087403,PhysRevB.95.235114}, theoretical~\cite{EPL.80.47003,PhysRevLett.101.106406, PhysRevB.75.214414,PhysRevB.77.134428,NJP.11.113038,LuoPhysRevB.89.165103,LuoPhysRevB.92.035109,PhysRevLett.109.117401,PhysRevLett.110.117005,PhysRevLett.106.157205,PhysRevB.86.125103}, and computational ~\cite{JohnstonPhysRevB.82.064513,NatCommun.4.1470,NJP.11.113038,PhysRevLett.106.157205,PhysRevB.83.245133,PhysRevLett.110.265502} approaches have investigated RIXS across a wide variety of systems in various dimensions for a range of elementary excitations at different X-ray edges. While the effect of magnon-magnon interactions on the indirect K-edge RIXS spectrum has been investigated both experimentally and theoretically in the antiferromagnetic (AF) and the collinear antiferromagnetic (CAF) phases~\cite{EPL.80.47003,PhysRevB.75.214414,PhysRevB.77.134428,NJP.11.113038,LuoPhysRevB.89.165103}, till date there is no study (theoretical or experimental), which \emph{exculsively} investigates the role of many body magnon-phonon interaction on the indirect  K-edge bimagnon RIXS intensity spectrum. At the K--edge spin angular momentum is conserved in the indirect RIXS process due to the lack of spin--orbit coupling in the 1s electron. Thus the double spin--flip bimagnon excitation is the leading process at the K--edge. For the higher angular momentum shells $(L, M, N, ...)$ with finite orbital momentum single spin flip excitations are allowed \cite{RevModPhys.83.705}.

As a probe RIXS has a high degree of senstivity towards the local environment. Thus it is a natural question to ask - How can phonons which produce lattice modulations (local vibrations) affect the magnetic RIXS spectrum? How can a multimagnon RIXS excitation spectrum, such as that of a bimagnon, be affected by phonons? The interplay of phonons with bimagnons offer the potential to uncover physical phenomena which has been overlooked till date. With next generation beamlines being constructed globally and experimental initiatives likely to probe phonon effects in correlated materials, answers to the above questions are imperative and timely. The theoretical study in this article offers insight on the key experimentally observable signatures which delineate magnon-magnon and magnon-phonon interaction effects in the indirect K-edge RIXS spectrum.

Past investigation on spatial anisotropy and significant frustration within the square lattice Heisenberg magnet has led to the prediction of a two-peak bimagnon structure \cite{LuoPhysRevB.89.165103}. The proposed two-peak structure was a consequence of the bimagnon spectrum's sensitivity to microscopic magnetic interactions. But, in real materials lattice vibrations do matter. Thus, a realistic theoretical model which provides a true depiction of the materials under investigation with a comprehensive account of magnon-phonon coupling is called for.

The coupling between magnetic and lattice degrees of freedom can generate novel physical phenomena. For example, it can effect electronic degrees of freedom \cite{MoserPhysRevLett.115.096404,AmentEPL.95.27008} induce multiferroic order \cite{Tothnatcomm13547}, create magnon-phonon excitation effects in Raman spectroscopy \cite{JohnstonPhysRevB.82.064513}, and have an effect on thermal conductivity \cite{ChernyshevPhysRevB.92.054409} and optical conductivity \cite{GruningerPhysRevB.62.12422}. Past theoretical studies on magnon-phonon interaction in quantum Heisenberg magnets have alluded to the fact that at low but finite temperatures phonons do play a role on influencing magnetic interaction. This fact has been especially studied within the context of experimental and theoretical studies of spin-phonon interaction on the Raman spectra of Heisenberg antiferromagnets. In these studies it was found that significant broadening effects dominate both the single magnon and two-magnon line shape. Theoretical calculations have attributed the anomalous broadening of the two-magnon spectrum in cuprates to phonon effects \cite{LorSawPhysRevLett.74.1867,LorSawPhysRevB.52.9576,DirkPhysRevB.49.12176,DirkPhysRevB.52.1025,DevereauxPhysRevB.51.505,JDLeeJPSJ.66.442}.

Unfortunately, Raman spectroscopy is a zero wave vector probe~\cite{RevModPhys.79.175}. Therefore it limits the amount of physical information that can be extracted. But, RIXS is \emph{not}. The high energy X--ray photons in the RIXS experiments allow for large transferred momenta, with the zero wavevector reproducing the Raman response. Thus, it is appropriate to consider RIXS to explore the full energy-wavevector range to study the effects of magnon-phonon coupling on the bimagnon excitation spectrum. Recently, there has been some experimental ~\cite{PengPhysRevB.92.064517,YavasJPCM22.485601,MoserPhysRevLett.115.096404} and theoretical ~\cite{AmentEPLphonon,DevereauxPhysRevX.6.041019,YavasJPCM22.485601} studies devoted to the study of electron-phonon coupling and its effect on the RIXS spectrum. We note, that our study is different from the existing ones since we are primarily concerned with the role of magnon-phonon interaction on the bimagnon excitation spectrum.

In this article, we compute the effects of magnon-phonon coupling on the indirect K-edge bimagnon resonant inelastic x-ray scattering (RIXS) intensity spectrum of a square lattice Heisenberg antiferromagnet. The magnetic model theoretically investigated includes both the nearest-neighbor (nn) and next-nearest neighbor (nnn) magnetic and magnon-phonon interaction effects. Since the Heisenberg magnetic model is considered upto the nnn interaction (see Eq.~\ref{eq:heis}), to be consistent in our theoretical formulation we include the magnon-phonon coupling beyond the nn interaction. Using Dyson-Maleev representation of spin operators, the Bethe-Salpeter ladder approximation scheme for the bimagnon interacting channel, and considering the lowest order magnon-phonon-magnon scattering interaction we elucidate the distinct features in the RIXS X-ray spectrum.

Our analysis of the RIXS spectrum, performed for a wide range of model parameters in both the antiferromagnetic and the collinear antiferromagnetic phases of this model, suggest several contrasting behavior in the antiferromagnetic and the collinear antiferromagnetic phase.  We consider damping effects in our calculation due to the presence of longitudinal acoustic phonons in our model. We find that in the AF phase the RIXS intensity spectrum attains a maximum value primarily localized around the K $\left(\pm\frac{\pi}{2}, \pm \frac{\pi}{2}\right)$ - point. Within the nearest-neighbor model the system is weakly sensitive to the presence of magnon-phonon interactions. For most parameter choices the feature developed is a shoulder in the RIXS spectrum. But, inclusion of the nnn magnetic and magnon-phonon coupling  within the isotropic model leads to a splitting of the peak. In contrast, for the CAF phase in the isotropic model the intensity is broadly distributed with a significant scattering intensity located around the Y $\left(\pm\frac{\pi}{2}, 0\right)$ - point. The rest of the spectral weight appears along the K to M $\left(\pm\pi, 0\right)$ path in the Brillouin zone (BZ). Furthermore, in the CAF phase for suitable anisotropy, nearest-, and next-nearest neighbor interaction parameters the phonon effects can manifest itself as a distinct peak both below and above the bimagnon peak. Such a feature is in contrast to the AF spectrum where the effect due to the phonon peak was located consistently beyond the bimagnon peak in the high energy end of the spectrum. Within the anisotropic model the AF RIXS spectrum is merely broadened without any special peak or shoulder development. However, in the CAF phase we find the RIXS bimagnon-phonon spectrum to be more sensitive to anisotropy compared to its antiferromagnetic counterpart. The final RIXS spectra is a result of intricate many body magnon-magnon interactions, influenced by the effect of many-body mangon-phonon interactions.

This article is organized as follows. In Sec.~\ref{Sec:spin-ham} we introduce the nn and nnn Heisenberg Hamiltonian including the effects of phonons. In Sec.~\ref{Sec:phononModel} we write down explicitly the contribution arising from the spin-phonon coupling. In Sec.~\ref{Sec:RIXS} we state the RIXS operator, the bimagnon Green function, the corresponding Bethe-Salpeter equation, and the phonon Green function contributions. In Sec.~\ref{Sec:Results} we present and discuss our results on the effect of damping and magnon-phonon coupling (Sec.~\ref{Subsec:Dampmagrixs}), phonon contribution to bimagnon RIXS spectrum (Sec.~\ref{Subsec:Phononrixs}), frustration and magnon-phonon coupling effects on the AF phase bimagnon phonon spectrum (Sec.~\ref{Subsec:Frustrixs}), CAF phase RIXS spectrum (Sec.~\ref{Subsec:caf}), and anisotropy effects in both the AF and CAF phase (Sec.~\ref{Subsec:anisfrusmagph}). Finally, in Sec.~\ref{Sec:Conclu} we provide our conclusions.

\begin{figure}[t]
\centering
{\subfigure[~Rigid Lattice]{
\includegraphics[scale=0.073]{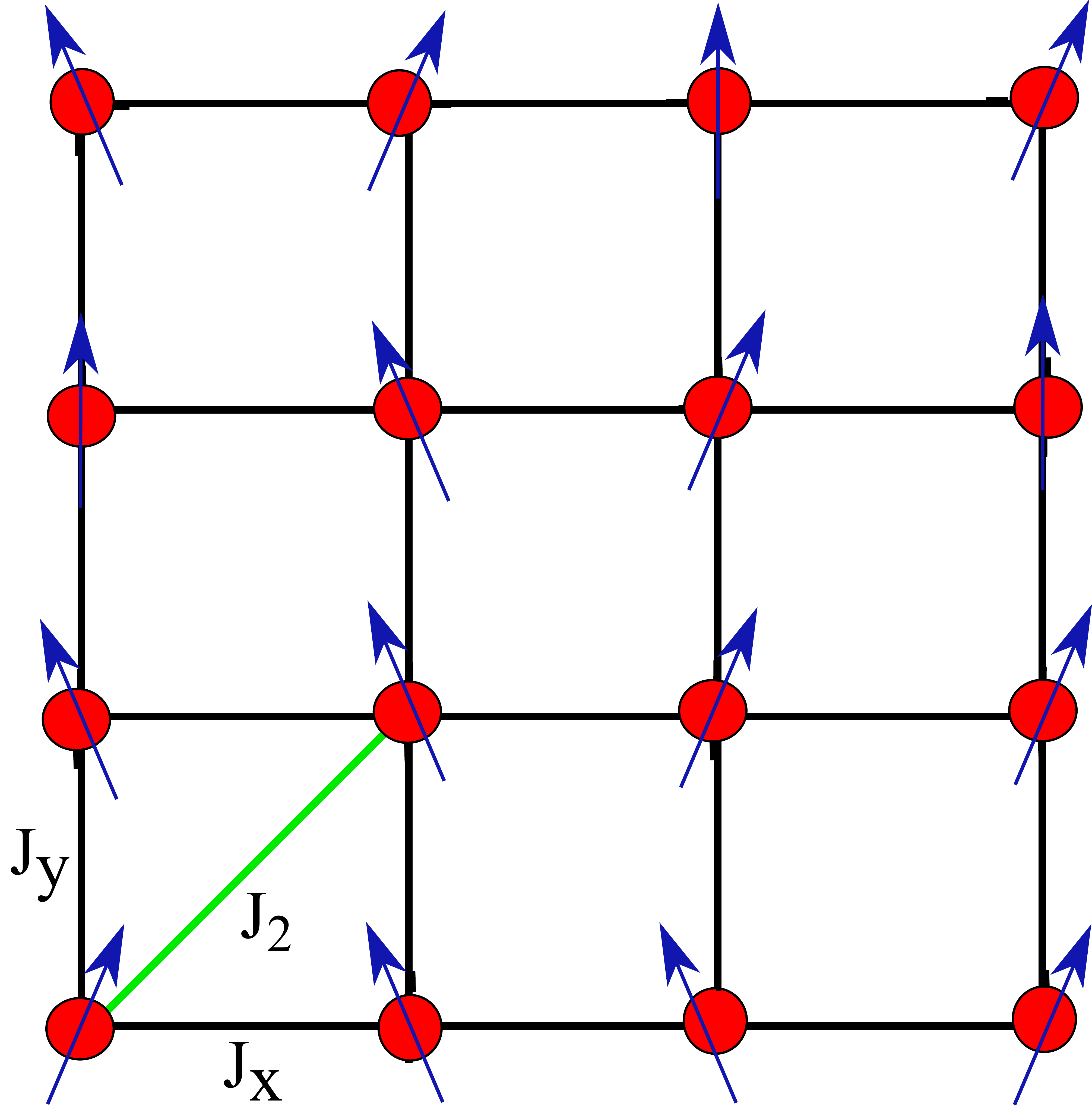}}\label{fig:1a}}
{\subfigure[~Deformabale lattice]{
\includegraphics[scale=0.025]{Fig1b.pdf}}\label{fig:1b}}
\caption{(Color online) Spatially anisotropic Heisenberg model on a square lattice with exchange interactions J$_{x,y}$ (nearest-neighbor along x,y bond) and  J$_{2}$ (next-nearest neighbor). (a) Lattice with rigid bonds. The exchange constants are independent of any spatial variation. (b) Lattice with deformable bonds modelled via spatially dependent exchange interactions, J$_{x,y}(\bold{r}_{ij})$ and J$_{2}(\bold{r}_{ij})$.}
\label{fig:fig1}
\end{figure}

%%%%%%%%%%%%%%%%%%%%%%%%%%%%%%%%%%%%%%%%%%%%%%%%%%%%%%%%%%%%%%%%
\section{Model}\label{Sec:Model}
%%%%%%%%%%%%%%%%%%%%%%%%%%%%%%%%%%%%%%%%%%%%%%%%%%%%%%%%%%%%%%%%
%%%%%%%%%%%%%%%%%%%%%%%%%%%%%%%%%%%%%%%%%%%%%%%%%%%%%%%%%%%%%%%%
\subsection{Heisenberg Hamiltonian}\label{Sec:spin-ham}
%%%%%%%%%%%%%%%%%%%%%%%%%%%%%%%%%%%%%%%%%%%%%%%%%%%%%%%%%%%%%%%%
The frustrated $J_1-J_2$ model~\cite{PhysRevB.78.052507,PhysRevB.79.092416,PhysRevB.84.155108}
and its spatially anisotropic $J_x-J_y-J_2$ version ~\cite{PhysRevB.82.144407,PhysRevB.81.024505,PhysRevB.81.165101} can support both
the $(\pi,\pi)$-AF and the $(\pi,0)$-CAF phase. There are several excellent material realizations of these models in cuprates \cite{ElbioRevModPhys.66.763},
pnictides \cite{DaiRevModPhys.87.855,StewartRevModPhys.83.1589}, and complex vanadium oxide compounds ~\cite{PhysRevB.79.214417}. For our purpose we generalize the model to include the effect of spin-lattice interaction by expanding the spatially dependent exchange interactions $J_{x,y}(\bold{r}_{ij})$ and $J_{2}(\bold{r}_{ij})$. The presence of lattice vibrations (phonons) cause the ionic distances to vary, which in turn are expected to induce magnon-phonon interactions ~\cite{DixonPhysRevB.21.2851}. We write the spin S=1/2 anisotropic $J_{x}-J_{y}-J_{2}$ Heisenberg model on a two dimensional square lattice as
\begin{equation}
\begin{aligned}\label{eq:heis}
 H=&\frac{1}{2}\sum_{i,\delta_{x}} J_{x}(\bold{r}_{i\,i+\delta_{x}})\bold{S}_{i} \cdot \bold{S}_{i+\delta_{x}} +\frac{1}{2}\sum_{i,\delta_{y}} J_{y}(\bold{r}_{i\,i+\delta_{y}})\bold{S}_{i} \cdot \bold{S}_{i+\delta_{y}}\\
 &+ \frac{1}{2}\sum_{i,\delta_{2}} J_{2}(\bold{r}_{i\,i+\delta_{2}})\bold{S}_{i} \cdot \bold{S}_{i+\delta_{2}},
\end{aligned}
\end{equation}
where $\bold{\delta_{2}}=\bold{\delta_{x}}+\bold{\delta_{y}}$. Assuming first order deviations from the lattice equilibrium positions, the super-exchange coupling $J(\bold{r}_{ij})$ can be expanded as~\cite{DirkPhysRevB.49.12176,JDLeeJPSJ.66.442}
\begin{equation}
J(\bold{r}_{ij})=J(\bold{R}_{ij})+(\bold{u}_{i}-\bold{u}_{j}) \cdot \nabla J(\bold{r})|_{\bold{r}=\bold{R}_{ij}},
\end{equation}
where $\bold{r}_{i}=\bold{R}_{i} + \bold{u}_{i}$, $\bold{R}_{i}$ is the equilibrium position of the site i, and $\bold{u}_{i}$ is the displacement operator due to the lattice vibration. Henceforth, we set $J_{x}(\bold{R}_{ij})=J_{x}$, $J_{y}(\bold{R}_{ij})=J_{y}$ and $J_{2}(\bold{R}_{ij})=J_{2}$. The phonon lattice deviations can be quantized resulting in a Hamiltonian which consists of a pure spin contribution H$^{s}$ and one with a magnon-phonon contribution H$^{sp}$. We next analyze the model within the standard Dyson-Maleev spin wave theory approach outlined in Appendix~\ref{app:dmftbv}. Note, the choice of representation Holstein-Primakoff or Dyson-Maleev does not affect the results of our paper. In the AF phase we have
\begin{equation}
\begin{aligned}\label{HS}
H^{s}=&E_{0}+E_{1}\sum_{\bold{k}}\kappa_{\bold{k}}(\epsilon_{\bold{k}}-1)\\
&+ E_{1}\sum_{\bold{k}}\kappa_{\bold{k}}\epsilon_{\bold{k}}(\alpha^{\dagger}_{\bold{k}}\alpha_{\bold{k}}+
\beta^{\dagger}_{-\bold{k}}\beta_{-\bold{k}})+H^{4}.
\end{aligned}
\end{equation}
In the above equation we have $E_{0}=-\frac{N}{2}J_{x}S^{2}z(1+\zeta)(1-\frac{2\eta}{1+\zeta})$, where N is the total number of sites and the coordination number $z=2$. We also introduce the interaction ratios $\zeta=J_{y}/J_{x}$ and $\eta=J_{2}/J_{x}$, and $E_{1}=J_{x}Sz(1+\zeta)$ in the AF phase. In addition, we define the following functions
\begin{eqnarray}
\gamma_{1}(\bold{k})=\frac{\cos{k_{x}} + \zeta \cos{k_{y}}}{1+\zeta},\quad
\gamma_{2}(\bold{k})=\cos{k_{x}} \cos{k_{y}},\\
\kappa_{\bold{k}}=1 - \frac{2\eta}{1+\zeta}(1-\gamma_{2}(\bold{k})),\,
\gamma_{\bold{k}}=\frac{\gamma_{1}(\bold{k})}{\kappa_{\bold{k}}}, \, \epsilon_{\bold{k}}=\sqrt{1-\gamma^{2}_{\bold{k}}}.
\end{eqnarray}
The higher order term $H^{4}$ is given by the expression
\begin{equation}
\begin{aligned}
\label{eq:qvert}
H^{4}=&\frac{E_{1}}{2S}\sum_{\bold{k}}[A_{\bold{k}}(\alpha^{\dagger}_{\bold{k}}\alpha_{\bold{k}}+\beta^{\dagger}_{-\bold{k}}\beta_{-\bold{k}})
+B_{\bold{k}}(\alpha^{\dagger}_{\bold{k}}\beta^{\dagger}_{-\bold{k}}+\alpha_{\bold{k}}\beta_{-\bold{k}})]\\
&+\frac{E_{1}}{SN}\sum_{1,2,3,4}\delta_{\bold{G}}(1+2-3-4)u_{1}u_{2}u_{3}u_{4}\\
&\times (V^{(4)}_{1234}\alpha^{\dagger}_{1}\beta^{\dagger}_{-4}\beta_{-2}\alpha_{3}+ \cdots),
\end{aligned}
\end{equation}
where the $u_{k}$ and $v_{k}$ coefficients arise in the Bogoliubov transformation with $v_{k}=-x_{k}u_{k}$. The momentum labels $k_{1},k_{2},\ldots$ are abbreviated as $1,2,\ldots$. We invoke the conservation of momentum rule upto the reciprocal lattice vector $\bold{G}$ with $\delta_{\bold{G}}(1+2-3-4)$. The constant and the quadratic terms in $H^{4}$ arising from normal ordering procedure are known as Oguchi corrections. The coefficients are
\begin{equation}
\begin{aligned}
A_{k}&=A_{1}\frac{1-\gamma_{1}(k)\gamma(k)}{\epsilon_{k}}-\Delta_{A_{1}}(\cos k_{x}-\cos k_{y})\frac{\gamma(k)}{\epsilon_{k}}\\
&\quad +A_{2}\frac{1-\gamma_{2}(k)}{\epsilon_{k}},\\
A_{1}&=\frac{2}{N}\sum_{p}\frac{\gamma_{1}(p)\gamma(p)+\epsilon_{p}-1}{\epsilon_{p}},\\
\Delta_{A_{1}}&=\frac{2}{N}\sum_{p}\frac{\zeta}{(1+\zeta)^{2}}(\cos p_{x}-\cos p_{y})\frac{\gamma(p)}{\epsilon_{p}},\\
A_{2}&=\frac{2\eta}{1+\zeta}\frac{2}{N}\sum_{p}\frac{1-\epsilon_{p}-\gamma_{2}(p)}{\epsilon_{p}}.
\end{aligned}
\end{equation}
Note, the $A_{k}$ equation is different from the previously reported expression~\cite{PhysRevB.46.10763, PhysRevB.72.014403,PhysRevB.82.144407,KMPhysRevB.85.144420}. We find an additional contribution $\Delta_{A_{1}}$ that was previously ignored (please see Appendix~\ref{app:dmftbv}). At the bimagnon RIXS spectrum level the presence or absence of this Oguchi correction term does not change the results quantitatively or qualitatively. However, with the inclusion of anisotropic interactions and magnon-phonon coupling it is important to consider an accurate expression. The AF phase magnon-phonon RIXS spectrum is mildly affected, but, the CAF phase spectrum is unaffected. The quartic interaction vertex  in Eq.~\ref{eq:qvert}, relevant to our calculation, is given by the expression
\begin{equation}
\begin{aligned}\label{vertex}
V^{(4)}_{1234}=&-[\gamma_{1}(3-2)x_{3}x_{4}+\gamma_{1}(4-2)+\gamma_{1}(3-1)x_{1}x_{2}x_{3}x_{4}\\
&+\gamma_{1}(4-1)x_{1}x_{2}-\gamma_{1}(2)x_{4}-\gamma_{1}(1)x_{1}x_{2}x_{4}\\
&-\gamma_{1}(3+4-2)x_{3}-\gamma_{1}(3+4-1)x_{1}x_{2}x_{3}]\\
&+\frac{2\eta}{2(1+\zeta)}[\gamma_{2}(4-2)+\gamma_{2}(4-1)+\gamma_{2}(3-2)\\
&+\gamma_{2}(3-1)-\gamma_{2}(2)-\gamma_{2}(1)-\gamma_{2}(3+4-2)\\
&-\gamma_{2}(3+4-1)](x_{2}x_{4}+\Phi_{G}\,x_{1}x_{3}).
\end{aligned}
\end{equation}
where $\Phi_{\bold{G}}=e^{iG_{x}}$. In the CAF phase, $H^{s}$ and $H^{4}$ have the same form as in the AF phase with the new redefined coefficients
\begin{equation}
\begin{aligned}
\gamma'_{1}(k)&=\frac{\cos k_{x}(1+2\eta \cos k_{y})}{1+2\eta},\quad \gamma'_{2}(k)=\cos k_{y},\\
\kappa'_{k}&=1-\frac{\zeta}{1+2\eta}(1-\gamma'_{2}(k)),\quad \gamma'_{\bold{k}}=\frac{\gamma'_{1}(\bold{k})}{\kappa'_{\bold{k}}},\\
A'_{k}&=A'_{1}\frac{1-\gamma'_{1}(k)\gamma'(k)}{\epsilon'_{k}}-\Delta'_{A_{1}}\cos k_{x}(1-\cos k_{y})\frac{\gamma'(k)}{\epsilon'_{k}}\\
&\quad +A'_{2}\frac{1-\gamma'_{2}(k)}{\epsilon'_{k}},\\
\Delta'_{A_{1}}&=\frac{2}{N}\sum_{p}\frac{2\eta}{(1+2\eta)^{2}}\cos p_{x}(1-\cos p_{y})\frac{\gamma'(p)}{\epsilon'_{p}}.
\end{aligned}
\end{equation}
The analytical expressions for $E'_{1},A'_{1},A'_{2}$, and $V'^{(4)}_{1234}$ can be obtained by using the replacement $\zeta \leftrightarrow 2\eta$. The same replacement will also generate the coefficients $\gamma'_{1}(k),
\gamma'_{2}(k),\kappa'_{k}$,and $\epsilon'_{k}$ in $A_{1},A_{2}$, and $V^{(4)}_{1234}$.
%%%%%%%%%%%%%%%%%%%%%%%%%%%%%%%%%%%%%%%%%%%%%%%%%%%%%%%%%%%%%%%%
\subsection{Magnon-phonon Hamiltonian}\label{Sec:phononModel}
%%%%%%%%%%%%%%%%%%%%%%%%%%%%%%%%%%%%%%%%%%%%%%%%%%%%%%%%%%%%%%%%
In this section we focus on the magnon-phonon Hamiltonian contribution $H^{sp}$, generalized to include both the effects of spatial anisotropy and further neighbor interactions. Introducing the Taylor expansion of the exchange coefficients as mentioned earlier, we can write down an expression for the magnon-phonon Hamiltonian in the AF phase as
\begin{equation}
\label{eq:hsph}
\begin{aligned}
H^{sp}&=\sum_{i,\delta_{x}}(\bold{u}_{i}-\bold{u}_{i+\delta_{x}}) \cdot \nabla J_{x}(\bold{r}_{i\,i+\delta_{x}})\bold{S}^{A}_{i} \cdot \bold{S}^{B}_{i+\delta_{x}} \\
&\quad +\sum_{i,\delta_{y}}(\bold{u}_{i}-\bold{u}_{i+\delta_{y}}) \cdot \nabla J_{y}(\bold{r}_{i\,i+\delta_{y}})\bold{S}^{A}_{i} \cdot \bold{S}^{B}_{i+\delta_{y}}\\
&\quad + \frac{1}{2}\sum_{i,\delta_{2}}[(\bold{u}_{i}-\bold{u}_{i+\delta_{2}}) \cdot \nabla J_{2}(\bold{r}_{i\,i+\delta_{2}})\bold{S}^{A}_{i} \cdot \bold{S}^{A}_{i+\delta_{2}} \\
&\quad +(\bold{u}_{i+\delta_{x}}-\bold{u}_{i+\delta_{x}+\delta_{2}}) \cdot \nabla J_{2}(\bold{r}_{i+\delta_{x}\,i+\delta_{2}})\bold{S}^{B}_{i+\delta_{x}}\cdot\bold{S}^{B}_{i+\delta_{x}+\delta_{2}}]
\end{aligned}
\end{equation}
where the quantized displacement operator $\bold{u}_{i}$ expression is given by
\begin{equation}
\bold{u}_{i}=\sum_{\bold{q},\lambda}\sqrt{\frac{\hbar}{2Nm\Omega_{\lambda}^{ph}(\bold{q})}}\bold{e}(\bold{q},\lambda)e^{-i\bold{q} \cdot \bold{R}_{i}}\varphi_{\bold{q}\lambda},
\end{equation}
where $\Omega_{\lambda}^{ph}(\bold{q})$ is the dispersion of phonon in branch $\lambda$, $\bold{e}(\bold{q},\lambda)$ is the phonon's polarization vector, $\varphi_{\bold{q}\lambda}$ is the phonon operator, and m is the reduced ionic mass. To recast Eq.~\ref{eq:hsph} into its spin wave version we use the standard Dyson-Maleev transformation, followed by a Fourier transformation, and a subsequent Bogoliubov transformation to obtain the bosonized magnon-phonon Hamiltonian expression as
\begin{equation}
\begin{aligned}\label{HSP}
H^{sp}=&S\frac{E_{1}}{\sqrt{N}}\sum_{\bold{k}_{1},\bold{k}_{2}}\sum_{\bold{q},\lambda}\delta_{\bold{G}}(\bold{k}_{1}-\bold{k}_{2}-\bold{q})
\varphi_{\bold{q}\lambda}\,g_{x}(\bold{q},\lambda)\\
&\times[A_{\lambda}(\bold{k}_{1},\bold{k}_{2},\bold{q})\alpha^{+}_{\bold{k}_{2}}\alpha_{\bold{k}_{1}}+
B_{\lambda}(\bold{k}_{1},\bold{k}_{2},\bold{q})\beta^{+}_{-\bold{k}_{1}}\beta_{-\bold{k}_{2}}\\
&\quad +C_{\lambda}(\bold{k}_{1},\bold{k}_{2},\bold{q})\alpha^{+}_{\bold{k}_{2}}\beta^{+}_{-\bold{k}_{1}}
+D_{\lambda}(\bold{k}_{1},\bold{k}_{2},\bold{q})\alpha_{\bold{k}_{1}}\beta_{-\bold{k}_{2}}].
\end{aligned}
\end{equation}
In the above we have introduced the notation
\begin{equation}
g_{x}(\bold{q},\lambda)=\frac{\left | \nabla J_{x} \right |}{E_{1}}\sqrt{\frac{\hbar}{2m\Omega_{\lambda}^{ph}(\bold{q})}}.
\end{equation}
The spin-phonon coupling coefficients $A_{\lambda}$,$B_{\lambda}$,$C_{\lambda}$, and $D_{\lambda}$ in Eq. \ref{HSP} are given in Appendix \ref{app:magph} for both the AF and the CAF phase. Furthermore, in the following discussion we introduce the magnon-phonon coupling ratios for anisotropy and nnn, respectively as
\begin{equation}
\Lambda_{\zeta}=g_{y}(\bold{q},\lambda)/g_{x}(\bold{q},\lambda),~~~~ \Lambda_{\eta}=g_{2}(\bold{q},\lambda)/g_{x}(\bold{q},\lambda).
\end{equation}
It is possible to estimate a value for the nn magnon-phonon coupling from experimental data~\cite{KnollPhysRevB.42.4842}.
For example, in cuprates such as RBa$_{2}$Cu$_{3}$O$_{6}$ (R = Eu, Y), the change in the exchange energy $\nabla J_{x}$ for the Cu-O bond ranges between 2500 - 6000 cm$^{-1}${\AA}$^{-1}$. The exchange constant itself varies between 800 - 1000 cm$^{-1}$ (96 - 120 meV). With $\hbar \Omega_{ph}$ in a 20 - 40 meV interval and the reduced mass of a Cu-O bond system equal to 2.13 $\times$ 10$^{-23}$ g, we find that g$_{x}$ lies between 0.076 - 0.41. For our calculations we have chosen g$_{x}=0.28$. Reliable experimental data or theoretical estimates on further neighbor magnon-phonnon couplings are either rare to find or difficult to obtain. Thus, for our purposes we make an educated guess of the physically reasonable ratios to simulate the RIXS spectrum. We hope this provides further motivation, both experimentally and theoretically, to investigate the physics of the further neighbor magnon-phonon coupled quantum magnet systems.
%%%%%%%%%%%%%%%%%%%%%%%%%%%%%%%%%%%%%%%%%%%%%%%%%%%%%%%%%%%%%%%%
\section{Bimagnon and Magnon-Phonon RIXS}\label{Sec:RIXS}
%%%%%%%%%%%%%%%%%%%%%%%%%%%%%%%%%%%%%%%%%%%%%%%%%%%%%%%%%%%%%%%%
In this section we present the expression for the bimagnon RIXS scattering operator generalized to the cases of spatially dependent exchange interaction. Utilizing the standard definition of the bimagnon RIXS operator, valid within the ultrashort core-hole lifetime (UCL) expansion ~\cite{vandenBrink20052145, EPL.73.121,PhysRevB.75.115118}, we  have
\begin{equation}
\hat{O}_{\bold{q}}=\sqrt{\frac{2}{N}}\sum_{ij}e^{i\bold{q} \cdot \bold{r}_{i}} J(\bold{r}_{ij})\bold{S}_{i}\cdot\bold{S}_{j},
\end{equation}
In its bosonized form the operator reads as
\begin{equation}
\hat{O}_{\bold{q}}=\sum_{\bold{k}}N(\bold{q},\bold{k})(\alpha^{\dagger}_{\bold{k}+\bold{q}}\beta^{\dagger}_{-\bold{k}}+\alpha_{\bold{k}}\beta_{-\bold{k}-\bold{q}})+\cdots,
\end{equation}
where we have
\begin{equation}
\begin{aligned}
N(\bold{q},\bold{k})=&E_{1}\{[1+\gamma_{1}(\bold{q})+\frac{2\eta}{1+\zeta}(\gamma_{2}(\bold{k}+\bold{q})+\gamma_{2}(\bold{k})-1\\
&-\gamma_{2}(\bold{q}))](u_{\bold{k}+\bold{q}}v_{\bold{k}}+u_{\bold{k}}v_{\bold{k}+\bold{q}})+(\gamma_{1}(\bold{k} +\bold{q})+\gamma_{1}(\bold{k}))\\
&\times (u_{\bold{k}+\bold{q}}u_{\bold{k}}+v_{\bold{k}}v_{\bold{k}+\bold{q}})\}.
\end{aligned}
\end{equation}
$N'(\bold{q},\bold{k})$ in the CAF phase can be obtained with the replacement $\zeta \leftrightarrow 2\eta$ along with the corresponding coefficients $\gamma'_{1}(k),\gamma'_{2}(k),u'_{k}$, and $v'_{k}$. In the following discussion, the energy is in units of $E_{1}$. The scattering intensity is given by
\begin{equation}
I(\bold{q},\omega)\propto \sum_{n}\left |\langle n| \hat{O}_{\bold{q}} |0\rangle \right |^{2} \delta(\omega -\omega_{n0}),
\end{equation}
where $| n\rangle$ represents the excited states in the RIXS intermediate process, and $| 0 \rangle$ is the ground state.
The Fourier transform of the zero temperature time-ordered Green's function is given by
\begin{equation}
iG(\bold{q},\omega)=\int^{\infty}_{0}dt\,e^{i\omega t}\langle 0|\mathcal{T}\hat{O}^{\dagger}_{\bold{q}}(t)\hat{O}_{\bold{q}}(0)|0\rangle.
\end{equation}
Then the scattering intensity can be expressed from the Green's function as
\begin{equation}
I(\bold{q},\omega)=-\frac{1}{\pi}\Im \text{m}\,G(\bold{q},\omega).
\end{equation}

\begin{figure}[t]
\centering
\includegraphics[scale=0.25]{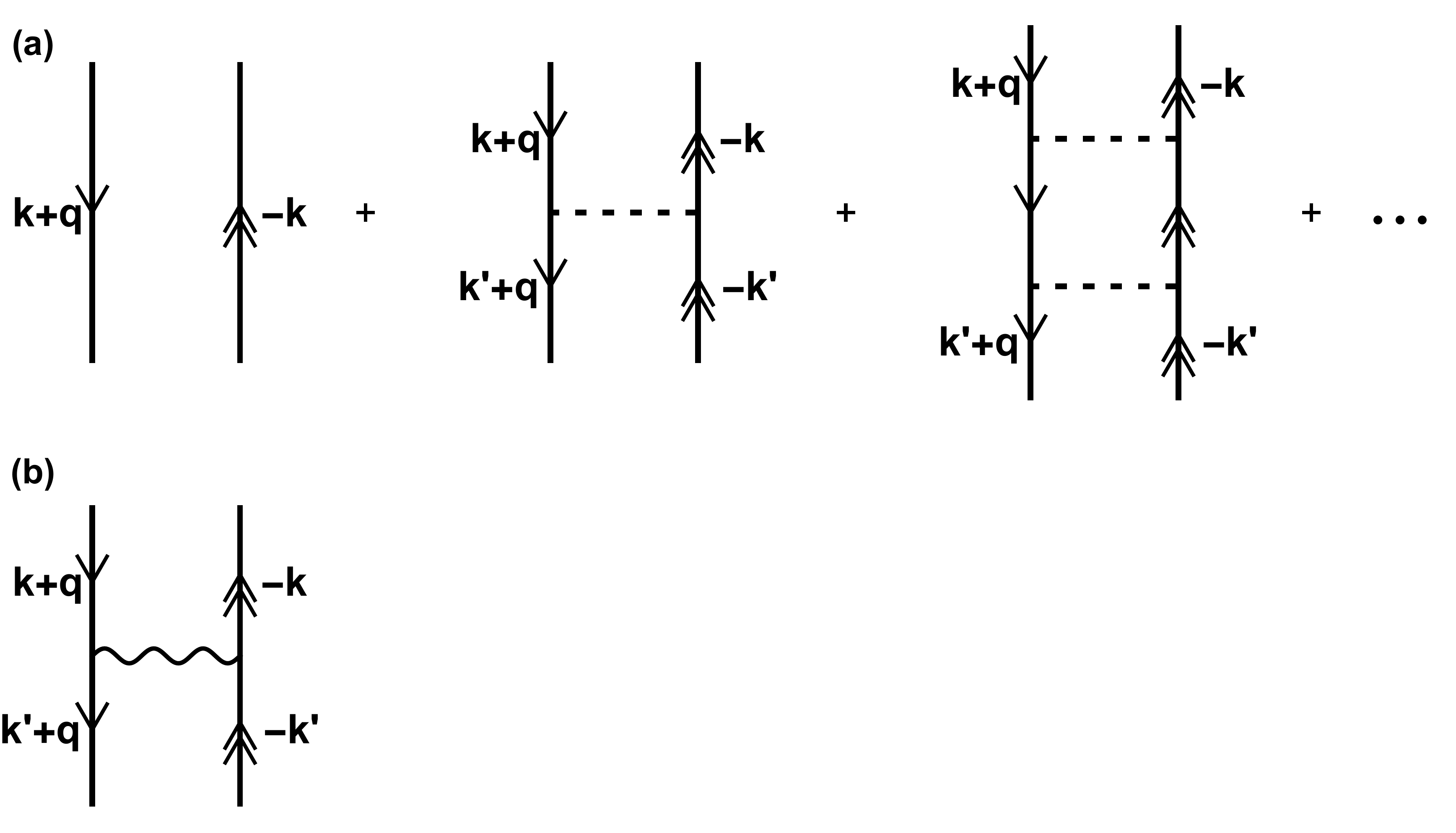}\label{fig:2}
\caption{Feynman diagrams used in the RIXS calculation. Single (double) arrows represent $\alpha \,(\beta)$ magnon. Dashed line represents magnon -- magnon interaction. Wiggly line represents magnon -- phonon interaction. {\bf k} and {\bf q} represent momentum and momentum transfer. (a) magnon -- magnon interaction vertices used in the ladder approximation Bethe--Salpeter scheme. (b) magnon -- phonon interaction vertex (truncated at the lowest order due to an approximation similar to Migdal's theorem). }
\label{fig:fig2}
\end{figure}

The perturbations for our problem are magnon--magnon interaction and magnon--phonon--magnon interaction. The Feynman diagrams are shown in Fig.~\ref{fig:fig2}. The Green's function $G(\bold{q},\omega)$ consists of a bimagnon part $G^{\text{bm}}(\bold{q},\omega)$ and a magnon-phonon-magnon part $G^{\text{m-ph-m}}(\bold{q},\omega)$. The bimagnon part is given by \cite{PhysRevB.75.214414, LuoPhysRevB.89.165103}
\begin{equation}
iG^{\text{bm}}(\bold{q},t)=\frac{2}{N}\sum_{\bold{k},\bold{k}'}N(\bold{q},\bold{k})N(\bold{q},\bold{k}')\Pi(\bold{q},t;\bold{k},\bold{k}'),
\end{equation}
where the interacting two magnon correlation function is defined as
\begin{equation}
i\Pi(\bold{q},t;\bold{k},\bold{k}')=\langle 0|\mathcal{T}\alpha_{\bold{k}+\bold{q}}(t)\beta_{-\bold{k}}(t)\alpha^{\dagger}_{\bold{k}'+\bold{q}}(0)\beta^{\dagger}_{-\bold{k}'}(0)|0\rangle.
\end{equation}
The bimagnon contribution include the effects of magnon-magnon interaction \cite{PowalskiPhysRevLett.115.207202,HamerPhysRevB.46.6276} can be be solved exactly using the Bethe-Salpeter equation. The interaction $u_{1}u_{2}u_{3}u_{4}V^{(4)}_{1234}$ in the spin Hamiltonian can be decomposed into 18 channels (see Appendix~\ref{app:vertgamma})
\begin{eqnarray}
&\frac{1}{S}u_{\bold{k}_{1}+\bold{q}}u_{\bold{k}}u_{\bold{k}+\bold{q}}u_{\bold{k}_{1}}V^{(4)}_{\bold{k}_{1}+\bold{q},\bold{k},\bold{k}+\bold{q},\bold{k}_{1}}
=\hat{\nu}(\bold{k})\hat{\Gamma}(\bold{q})\hat{\nu}^{T}(\bold{k}_{1}),\label{decom}
\end{eqnarray}
where $\hat{\nu}(\bold{k})$ has dimensions of $1\times 18$, $\hat{\Gamma}$ is a matrix of dimension $18 \times 18$. The expressions for each of these quantities in the AF and CAF phase are given in Appendix~\ref{app:vertgamma}. After summing the ladder diagrams exactly
\cite{LuoPhysRevB.89.165103,PhysRevB.4.992,PhysRevB.45.7127,PhysRevB.75.214414}, the two magnon Green's function can be expressed as a combination of several matrix products as
\begin{eqnarray}
G^{\text{bm}}(\bold{q},\omega) &=& G_{0}(\bold{q},\omega)\\ \non
&+&\hat{\mathcal{G}}(\bold{q},\omega)\hat{\Gamma}(q)[\hat{1}-\hat{R}(\bold{q},\omega)\hat{\Gamma}(q)]^{-1}\hat{\mathcal{G}}^{T}(\bold{q},\omega),
\end{eqnarray}
where $\hat{1}$ is a unit matrix of dimensions $18\times 18$ , and we define the non-interacting Green function and the non-interacting polarization propagtor, respectively, as
\begin{eqnarray}
G_{0}(\bold{q},\omega)&=&\frac{2}{N}\sum_{\bold{k}}N(\bold{q},\bold{k})^{2}\Pi_{0}(\bold{q},\omega;\bold{k}),\\
\Pi_{0}(\bold{q},\omega;\bold{k})&=&(\omega-\omega_{\bold{k}+\bold{q}}-\omega_{\bold{k}}+i0^{+})^{-1},
\end{eqnarray}
and
\begin{eqnarray}
&\hat{\mathcal{G}}(\bold{q},\omega)=\frac{2}{N}\sum_{\bold{k}}N(\bold{q},\bold{k})\Pi_{0}(\bold{q},\omega;\bold{k})\hat{\nu}(\bold{k}),\\
&\hat{R}(\bold{q},\omega)=\frac{2}{N}\sum_{\bold{k}}\Pi_{0}(\bold{q},\omega;\bold{k})\hat{\nu}^{T}(\bold{k})\hat{\nu}(\bold{k}).
\end{eqnarray}
The leading order of the magnon-phonon-magnon part (the zeroth order part is already included in bimagnon part), $G^{\text{m-ph-m}}$ takes the form
\begin{equation}
\begin{aligned}
\label{eq:gmph}
&G^{\text{m-ph-m}}(\bold{q},\omega)=\frac{2}{N^{2}}S^{2} \\
&\times \sum_{\bold{k},\bold{k}',\lambda}N(\bold{q},\bold{k})N(\bold{q},\bold{k}')g_{x}^{2}(\bold{k}-\bold{k}',\lambda)\\
&\times A_{\lambda}(\bold{k}+\bold{q},\bold{k}'+\bold{q},\bold{k}-\bold{k}')B^{*}_{\lambda}(\bold{k}',\bold{k},\bold{k}'-\bold{k})\\
&\times\frac{2}{\omega-\omega_{\bold{k}}-\omega_{\bold{k}'+\bold{q}}-\Omega^{ph}_{\lambda}(\bold{k}-\bold{k}')+i\Delta_{\bold{k}}+i\Delta_{\bold{k}'+\bold{q}}}\\
&\times\frac{1}{\omega-\omega_{\bold{k}}-\omega_{\bold{k}+\bold{q}}+i\Delta_{\bold{k}}+i\Delta_{\bold{k}+\bold{q}}}\\
&\times\frac{1}{\omega-\omega_{\bold{k}'}-\omega_{\bold{k}'+\bold{q}}+i\Delta_{\bold{k}'}+i\Delta_{\bold{k}'+\bold{q}}}.
\end{aligned}
\end{equation}
In the above Green functions we have introduced the variable $\Delta_{\bold{k}}$ as magnon damping due to magnon-phonon-magnon interaction. We use the energy dispersion $\omega_{k}=\kappa_{k}\epsilon_{k}+A_{k}/2S$ in units of $E_{1}$.
%%%%%%%%%%%%%%%%%%%%%%%%%%%%%%%%%%%%%%%%%%%%%%%%%%%%%%%%%%%%%%%%
\section{Bimagnon RIXS Spectra Results}\label{Sec:Results}
In this section we systematically investigate the effect of damping, magnon-phonon interaction, magnon-magnon interaction,
and anisotropy on the K-edge bimagnon phonon indirect RIXS spectrum. We compute the RIXS spectrum along the BZ traversing the path $\Gamma$:(0,0) $\rightarrow$ K$:\left(\frac{\pi}{2}, \frac{\pi}{2}\right)$ $\rightarrow$  M:$\left(\pi, 0 \right)$ $\rightarrow$ $\Gamma$:(0,0).

{\subsection{Damping and magnon-phonon coupling effects}\label{Subsec:Dampmagrixs}}
We consider longitudinally polarized acoustic phonons, $\bold{e}(\qq,\lambda)\parallel \qq$, with a dispersion given by ~\cite{DirkPhysRevB.49.12176}
\begin{equation}
\Omega_{\lambda}^{ph}(\qq)=\Omega^{ph}m(\qq),
\end{equation}
where
\begin{equation}
m(\qq)=\sqrt{\sin^{2}(q_{x}/2)+\sin^{2}(q_{y}/2)}.
\end{equation}
%We also introduce
%\begin{equation}
%g_{x}=\frac{|\nabla J_{x}|}{E_{1}}\sqrt{\frac{\hbar}{2m\Omega^{ph}}},
%\end{equation}
To keep the discussion and analysis of the model tractable we first investigate the isotropic version of the model. In spatially isotropic limit of the model, $\zeta=1$,  we set $J_{x}=J_{y}=J_{1}$ and $g_{x}=g_{y}=g_{1}$. Thus, the overall energy scale is given by $E_{1}=2J_{1}Sz$. The choice of $(\zeta,\eta)$ parameters are dictated by the magnetization phase diagram of the $J_x-J_y-J_2$ model ~\cite{PhysRevB.82.144407}.
\begin{figure}[t]
\centering
\includegraphics[scale=0.2]{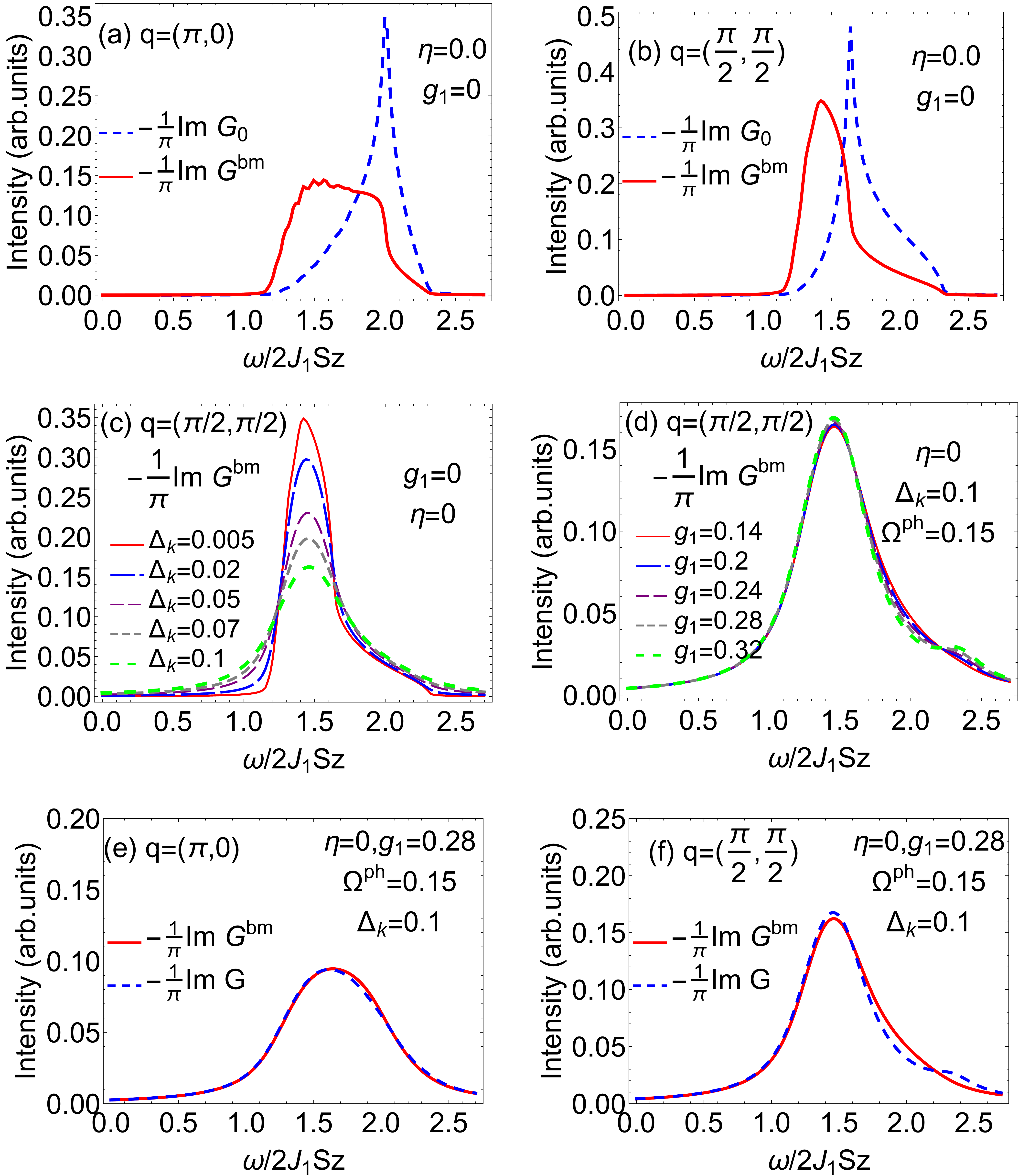}
\caption{(Color online) AF phase (isotropic nearest neighbor case): (a) - (b) Undamped non--interacting (G$_{0}$) and interacting bimagnon (G$^{\text{bm}}$) indirect RIXS intensity spectrum plots at the M- and K- points in BZ. (c) Effect of damping on the bimagnon RIXS plot for a range of damping parameter $\Delta_{k}$ values computed at the K- point in BZ. (d) Effect of nearest-neighbor magnon-phonon coupling $g_{1}$ at fixed damping parameter value of $\Delta_{k} = 0.15 E_{1}$ at the K- point in BZ. (e) - (f) Combined effect of the nearest-neighbor phonon contribution $g_{1}$ and damping on the bimagnon RIXS spectrum at both the  M- and K- points. Note, the presence of magnon-phonon coupling \emph{necessitates} the inclusion of damping in the RIXS calculation. $\Omega^{ph}=0.15 E_{1}$ is used in (d)-(f). $G$ refers to total, please see Sec.\ref{Sec:RIXS} for definition of G.}
\label{fig:fig3}
\end{figure}

In the absence of a magnetic field magnons in a 2D square lattice are not damped~\cite{KopietzPhysRevB.41.9228,HalperinPhysRevB.42.2096}. However, with the inclusion of magnon-phonon interaction the lifetime of the magnons are affected~\cite{DirkPhysRevB.49.12176,DirkPhysRevB.52.1025}. In the regime where the sound velocity is less than the magnon velocity, there exists spontaneously occuring decay processes where a magnon can decompose into a magnon and a phonon. Even at zero temperature, quantum fluctuations arising from magnon-phonon interactions can damp spin wave excitations. Therefore, presence of spin-lattice couplings can have an effect on the RIXS spectrum. Thus, a proper treatment of the magnon-phonon interaction should consider the damping effect.

\begin{figure}[t]
\centering
\includegraphics[scale=0.2]{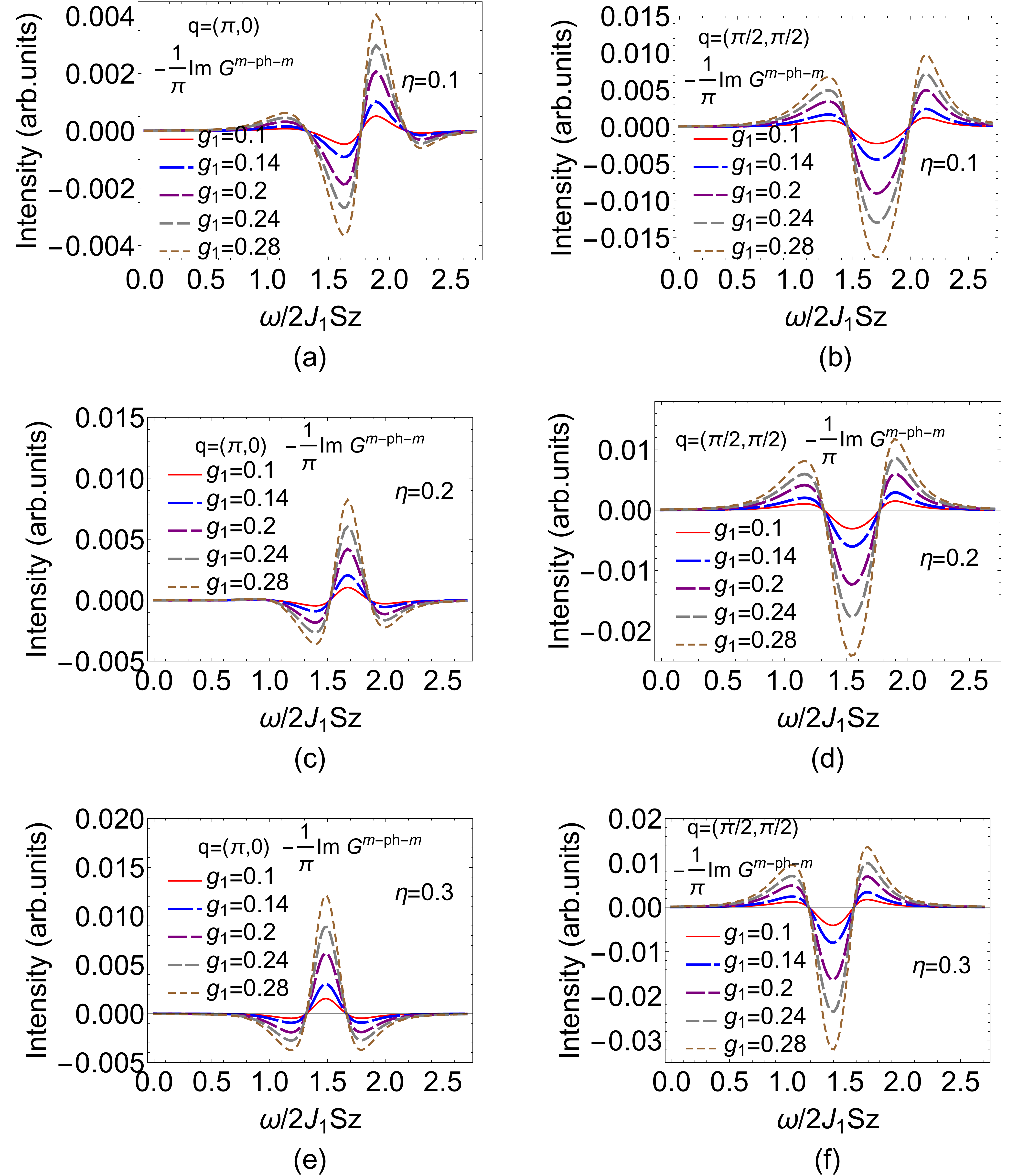}\label{fig:4a}
\caption{(Color online) Magnon-phonon-magnon intensity line spectrum comparison at the M- and
K- points in BZ for varying $g_{1}$ and $\eta$ values in the AF phase. $\Lambda_{\eta} = 0$ , $\Delta_{k}=0.1 E_{1}$, and $\Omega^{ph}=0.15 E_{1}$. $G^{\text{m-ph-m}}$ refers to magnon-phonon-magnon Greens function (see Eq.~\ref{eq:gmph}).}
\label{fig:fig4}
\end{figure}

In Fig.~\ref{fig:fig3} we display our calculations of the energy renormalization and damping effects within our model. In Figs.~\ref{fig:fig3}(a) -~\ref{fig:fig3}(b) we display the undamped RIXS intensity spectrum at the $M$- and the $K$- point of the BZ in the AF phase. Note, the presence of the van-Hove singularity like sharp peaks in the non-interacting case (dashed lines). These singular structures disappear when we include the two-magnon interaction (red solid lines) \cite{PhysRevB.75.214414}. The scattering of two magnons indeed changes the structure of the response function implying that it is no longer a simple product of the RIXS matrix element and the density of states.

In Fig.~\ref{fig:fig3}(c) we study damping on the RIXS spectrum. With increasing damping strength the spectrum is broadened, more so in the low energy regime where the spectrum height decreases with increasing damping strength. The high energy tail of the intensity pattern is not much affected. While damping could potentially arise from various microscopic mechanisms within a 2D square lattice problem, in this article we mainly focus on the effect of magnon-phonon interaction. Also for simplicity, we consider a phenomenological phonon induced constant damping to describe the imaginary part of the self energy. Introducing a constant damping to describe phonon effects is inspired by some previous theories of two magnon Raman spectra in cuprates~\cite{WeberPhysRevB.40.6890,KnollPhysRevB.42.4842}. Thus, in our calculations we set the damping parameter $\Delta_{\kk}=0.1E_{1}$. In Fig.~\ref{fig:fig3}(d) we display the trend on the RIXS spectrum that would arise when the nn magnon-phonon coupling $g_{1}$ is increased in strength. Clearly, beyond a threshold magnon-phonon coupling strength a shoulder peak develops. However, the mere inclusion of the magnon-phonon coupling $g_{1}$ is not strong enough to induce any further features. In Figs.~\ref{fig:fig3}(e) -~\ref{fig:fig3}(f) we display the combined effects of damping and magnon-phonon coupling. At this stage, the nnn interaction is set to zero.

\begin{figure}[t]
\centering
{\subfigure{
\includegraphics[scale=0.17]{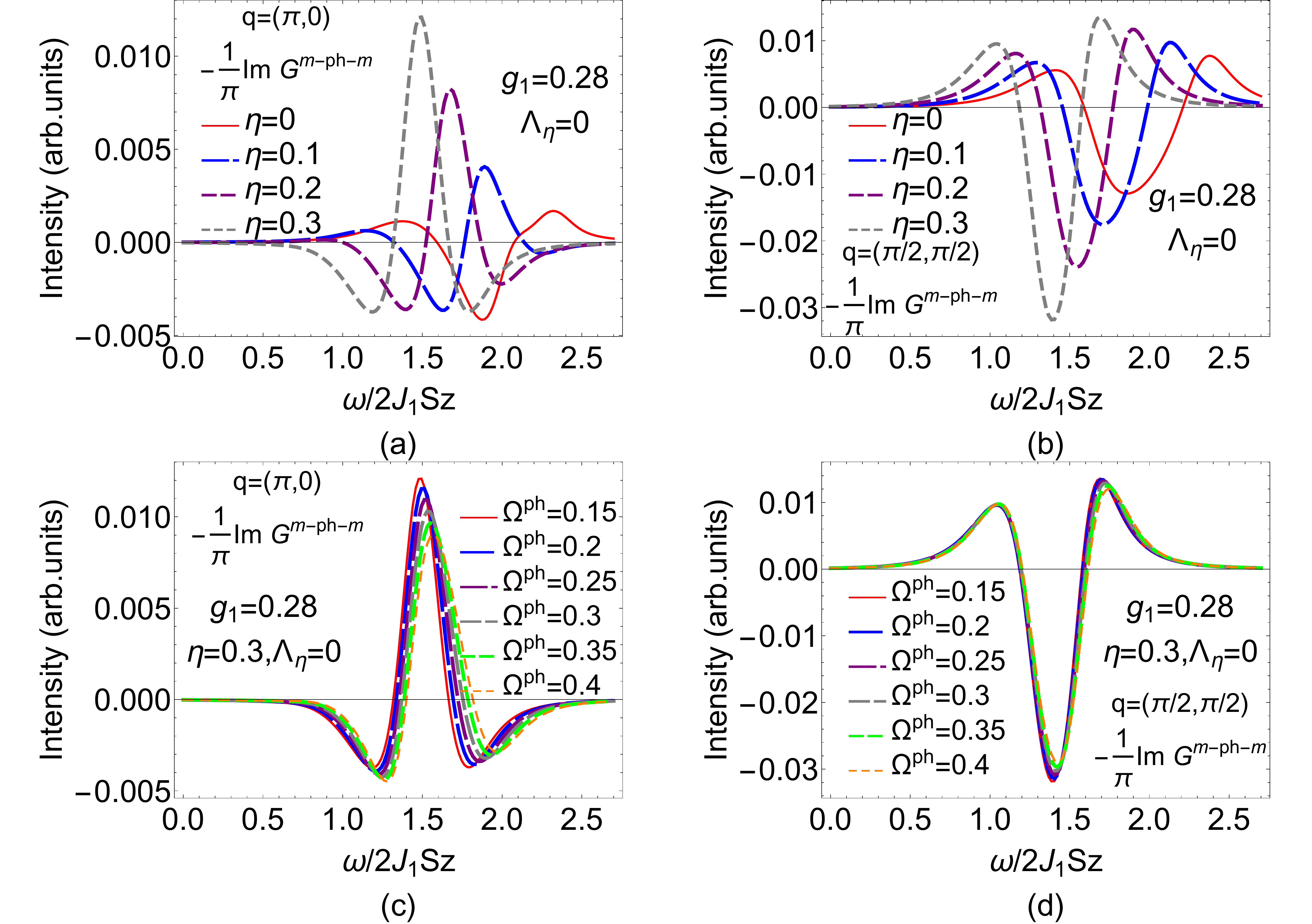}}\label{fig:5a}}
{\subfigure{
\includegraphics[scale=0.23]{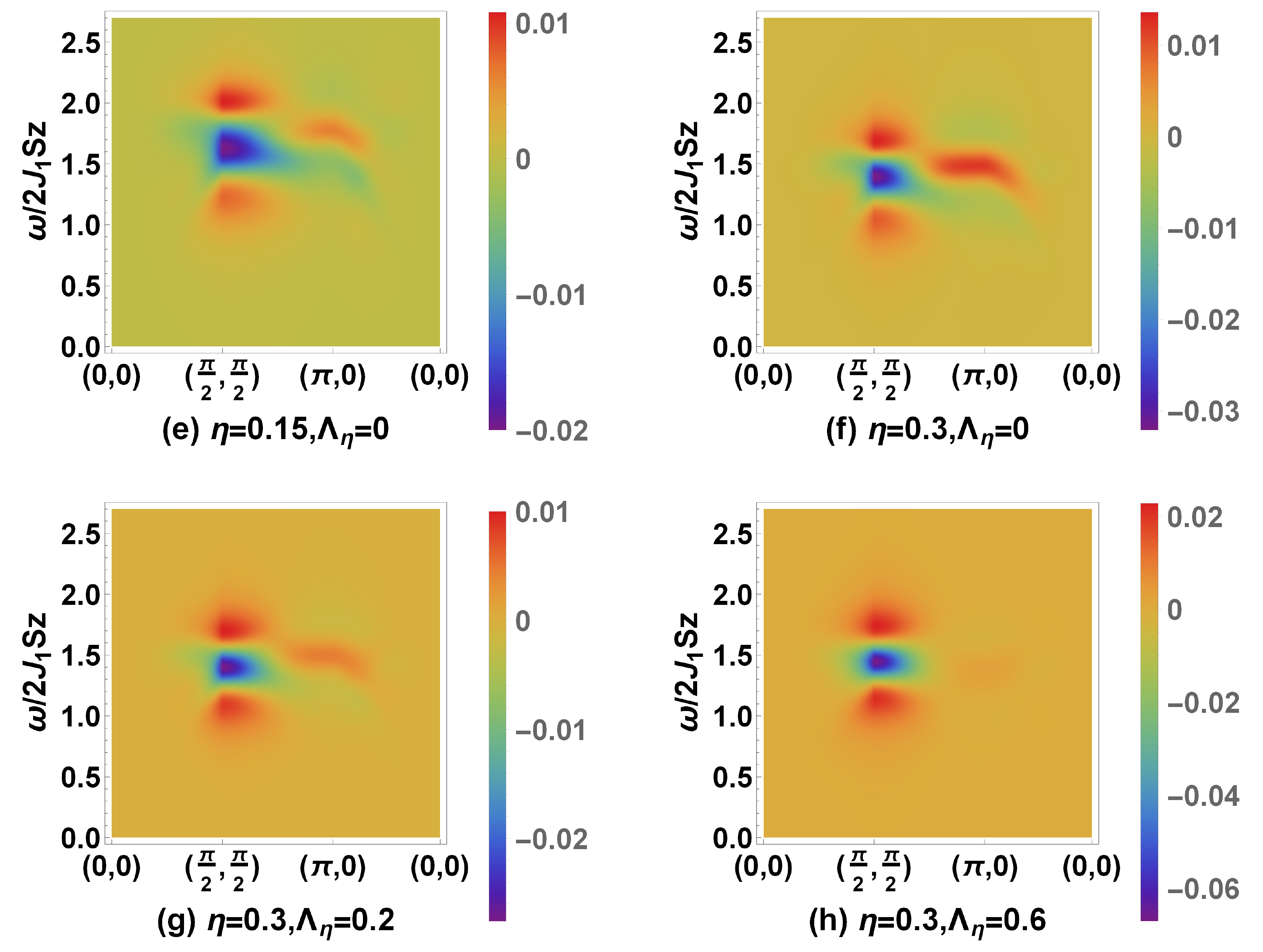}}\label{fig:5b}}
\caption{(Color online) AF phase: (a) - (b) Magnon-phonon-magnon intensity line spectrum computed at the M- and
K- points in BZ. (c) - (d) Effect of $\Omega^{ph}$ in magnon-phonon-magnon intensity at M point and K point. (e) - (h) 2D magnon-phonon-magnon RIXS intensity plot across the BZ for a range of frequency values computed with $g_{1} = 0.28$. The maximum contribution is localized around the K$:\left(\frac{\pi}{2}, \frac{\pi}{2}\right)$ - point. With increasing frustration the spectrum contribution enhances in strength around the BZ edge.  (g) - (h) Variation with respect $\Lambda_{\eta} = 0.2, 0.6$. Damping parameter fixed to a value of $\Delta_{k} = 0.1 E_{1}$ in all plots. $\Omega^{\text{ph}}=0.15 E_{1}$  in (a) - (b) and (e) - (h).}
\label{fig:fig5}
\end{figure}

{\subsection{Phonon contribution to RIXS spectrum}\label{Subsec:Phononrixs}}
The origin of magnon-phonon interaction is dynamical in nature. Thus summing up an infinite set of diagrams, as in
the bimagnon RIXS intensity case, can pose a serious calculation challenge. To proceed with the calculation  we note
that higher order terms generated by the magnon-phonon interaction decrease rapidly. Thus, akin to the celebrated Migdal
theorem used within the context of electron-phonon scattering in superconductivity we consider only the leading order magnon-phonon diagram, see Fig. \ref{fig:fig2}(b).

\begin{figure}[t]
\centering
\includegraphics[scale=0.25]{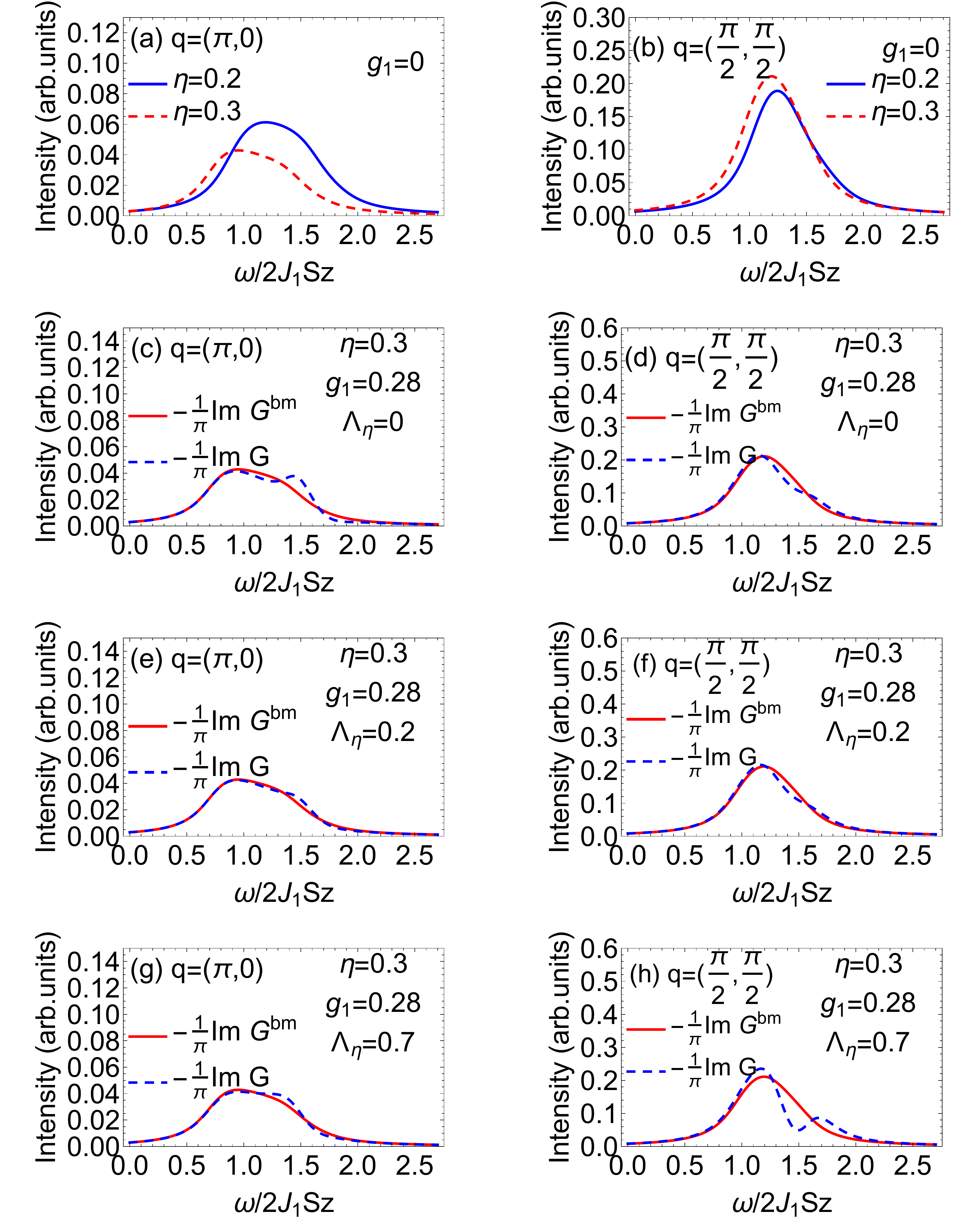}
\caption{(Color online) AF phase: (a) - (b) Interacting bimagnon RIXS intensity with constant damping $\Delta_{k}=0.1 E_{1}$. (c) - (h) Total (interacting bimagnon plus magnon-phonon) indirect RIXS intensity spectrum plots for various next-nearest neighbor ($\eta$), phonon nearest ($g_{1}$), and phonon next-nearest neighbor ($\Lambda_{\eta}$)
interaction parameters. We use $\Omega^{ph}=0.15 E_{1}$ in the plots (c) - (h). See Sec.~\ref{Sec:RIXS} for definitions of G and G$^{\text{bm}}$.}
\label{fig:fig6}
\end{figure}

In Fig.~\ref{fig:fig4} we showcase the effects of $g_{1}$ on the magnon-phonon-magnon intensity as the nnn frustration parameter is varied from 0.1 to 0.3.
It appears that for a given nnn interaction strength, wthin the isotropic model, the magnon-phonon interaction causes a rearrangement of the spectral strength. While at the $M-$ point the reshuffling of the spectral weight is quite prominent, that at the $K$- point is minimally affected. However, for both the locations in the BZ the spectral intensity scales in proportion to the magnon phonon coupling.

\begin{figure}[t]
\centering
\includegraphics[scale=0.25]{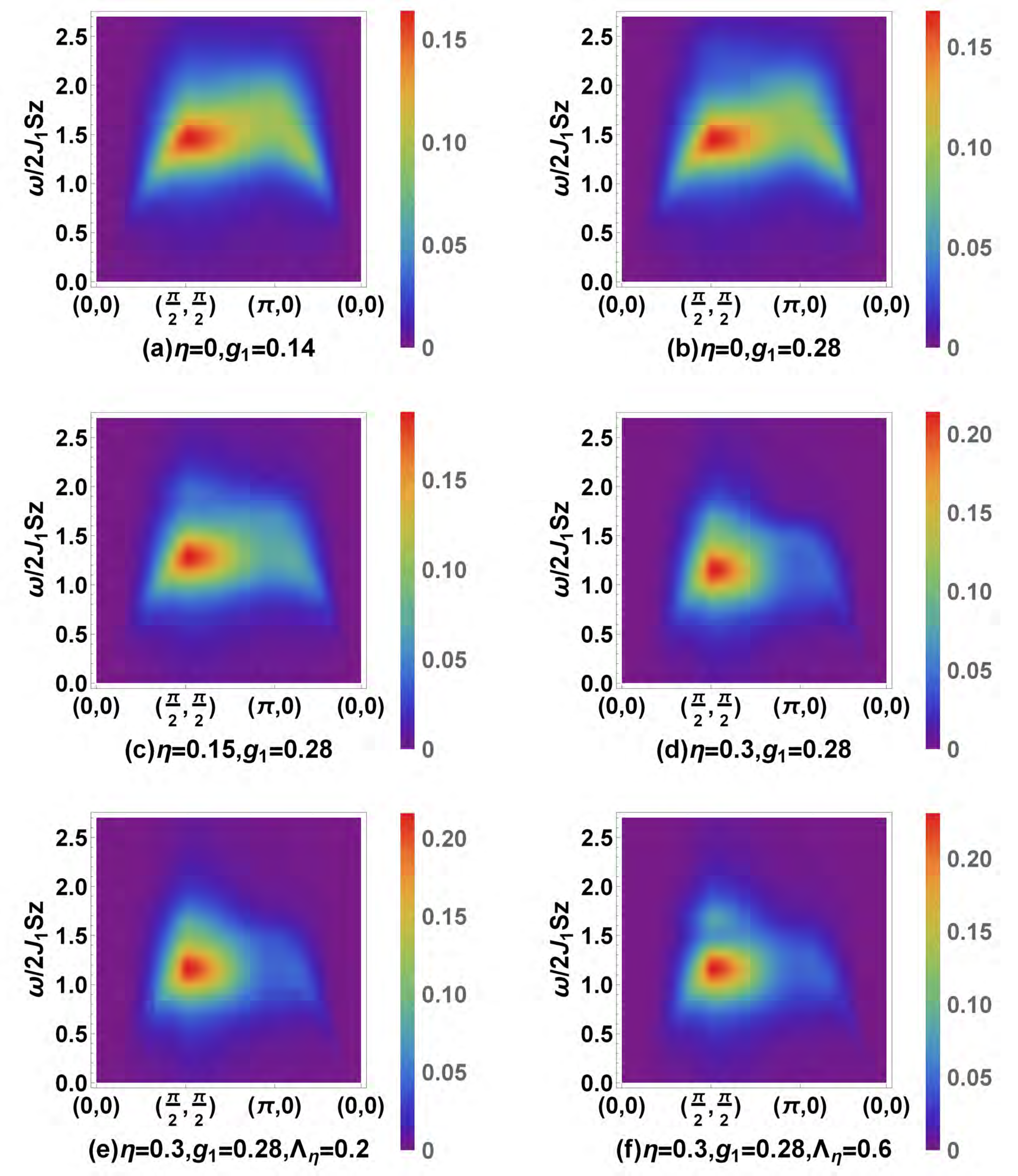}
\label{fig:7}
\caption{(Color online) AF phase, 2D total (interacting bimagnon plus magnon-phonon) indirect RIXS intensity spectrum plots for various
next-nearest neighbor ($\eta$), phonon nearest neighbor coupling ($g_{1}$), and phonon next-nearest neighbor ($\Lambda_{\eta}$)
interaction parameters across the entire BZ. Damping has been set to $\Delta_{k}=0.1 E_{1}$.}
\label{fig:fig7}
\end{figure}

In Figs.~\ref{fig:fig5}(a) and ~\ref{fig:fig5}(b) we investigated the magnon-phonon-magnon RIXS intensity spectrum as a function of $\eta$. It is observed that with increasing interaction the spectra has a downshift. Note, the miniscule unphysical negative contributions in the intensity is an artifact of the exclusion of the zeroth order contribution from the magnon-phonon-magnon vertex function~\cite{JDLeeJPSJ.66.442}. The final total RIXS intensity which includes the contribution from the phonon induced ladder interactions restores this term and naturally yields an overall physical positive RIXS intensity spectrum (see Fig.~\ref{fig:fig6} and Fig.~\ref{fig:fig7}). The observed spectral downshift is a characteristic feature that is also noticed in neutron ~\cite{PhysRevLett.86.5377,nature02576} and Raman spectrum ~\cite{ChenPhysRevLett.106.067002} with frustrated interactions. Our present calculations confirm that such an effect can also occur even within the magnon-phonon-magnon channel. Interestingly, while the trend towards downshift itself is robust, the peak-dip-peak structure appears to be dependent on the BZ location. At the characteristic $(\pi,0)$ (M point in BZ) the intensity pattern is complimentary to that observed at the K-point. In Figs.~\ref{fig:fig5}(c) and ~\ref{fig:fig5}(d) the effect of $\Omega^{ph}$ on the magnon-phonon-magnon intensity at M point and K point is displayed to give a sense of how the RIXS spectrum may be affected. We observe within our choice of parameters  $\Omega^{ph}$ does not affect the intensity spectrum.

From Figs.~\ref{fig:fig5}(e) -~\ref{fig:fig5}(f) it is clearly evident that with increasing nnn interaction $\eta$ the phonon RIXS spectrum undergoes a shift in the spectral weight. While most of the weight is localized around the K point, with enhanced nnn interaction some of the weight disperses towards the BZ corner. The 2D plot suggests that the phonon effects are primarly localized along the $(\pi/2,\pi/2) \rightarrow (\pi,0)$ line in BZ. However, with the inclusion of the nnn phonon interaction the spectral leaking is subdued, with the spectrum becoming more localized at the $K$ point, see Figs.~\ref{fig:fig5}(g) -~\ref{fig:fig5}(h).

{\subsection{Effect of frustration and magnon-phonon coupling}\label{Subsec:Frustrixs}}
In Fig.~\ref{fig:fig6} we show the combined effect of including nnn interaction and phonon coupling in the calculation.
From Fig.~\ref{fig:fig6}(a) -~\ref{fig:fig6}(b) it is clearly evident that nnn interaction introduces a downshift of the spectral weight. In Figs.~\ref{fig:fig6}(c) -~\ref{fig:fig6}(d) we show the dependence on the RIXS spectrum as the magnon--phonon coupling is introduced. Comparing the line plots for the $\eta=0.3$ case with that in Figs.~\ref{fig:fig6}(c) -~\ref{fig:fig6}(d) we notice that including magnon-phonon coupling causes a peak development, in addition, to introducing further broadening. The broadening effect is more prominent at the M- point. We notice from Figs.~\ref{fig:fig6}(e) -~\ref{fig:fig6}(h) that the effect of next-nearest neighbor magnon-phonon coupling is minimal at the M point, however, has a significant effect at the K- point. Further pronounced effects of the nnn magnon-phonon coupling contribution is clearly visible in Fig.~\ref{fig:fig6}(h). A 2D RIXS intensity pattern tracking the evolution of the total interacting RIXS peak development is shown in Fig.~\ref{fig:fig7}. We observe that the most prominent signal is at the K- point with satellite peaks due
to magnon-phonon coupling developing as the strength of the nnn contribution is enhanced. The effect of the nn magnon-phonon coupling appears to be minimal within this model. This fact is evident by observing the relatively unchanged RIXS spectrum features when the magnon-phonon coupling is changed.

\subsection{RIXS intensity in CAF phase}\label{Subsec:caf}
\begin{figure}[t]
\centering
\centering
{\subfigure{
\includegraphics[scale=0.5]{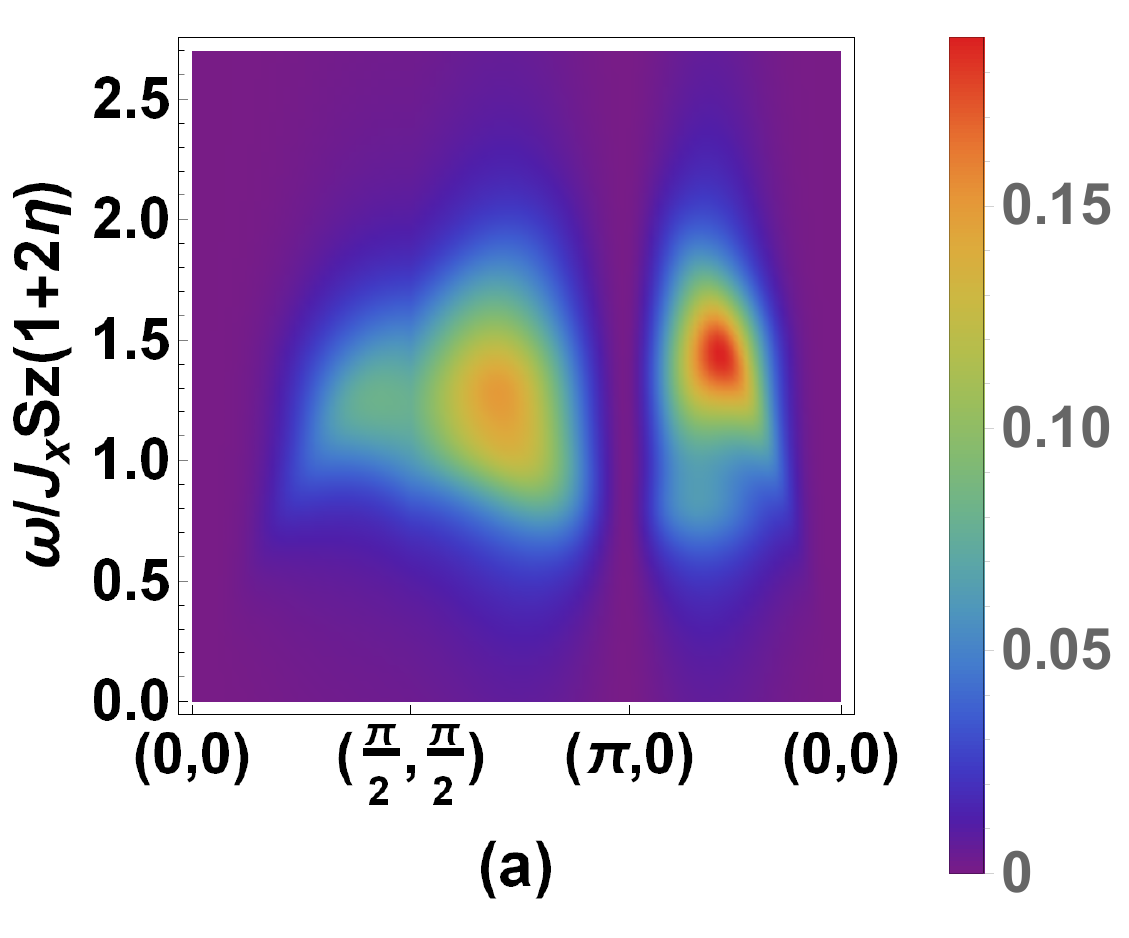}}\label{fig:8a}}
{\subfigure{
\includegraphics[scale=0.5]{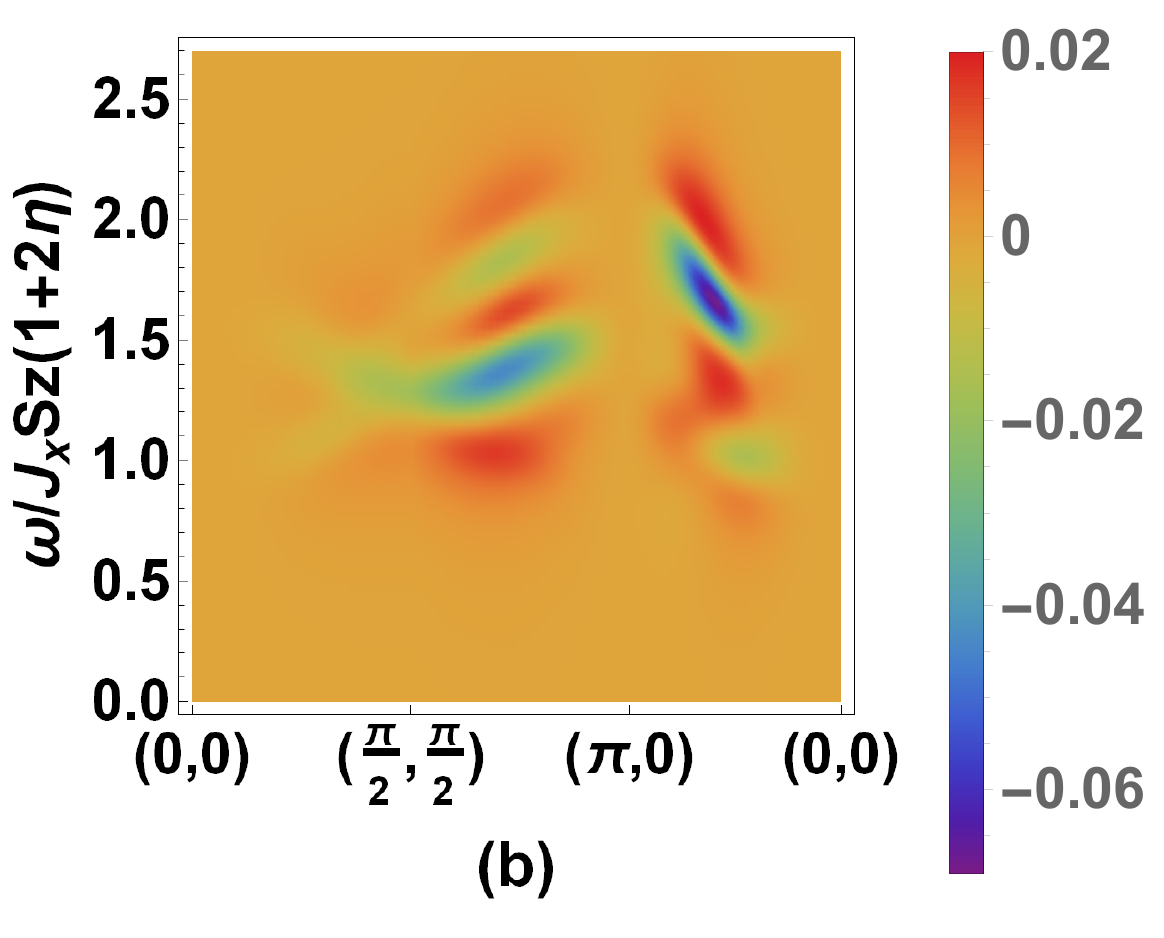}}\label{fig:8b}}
{\subfigure{
\includegraphics[scale=0.5]{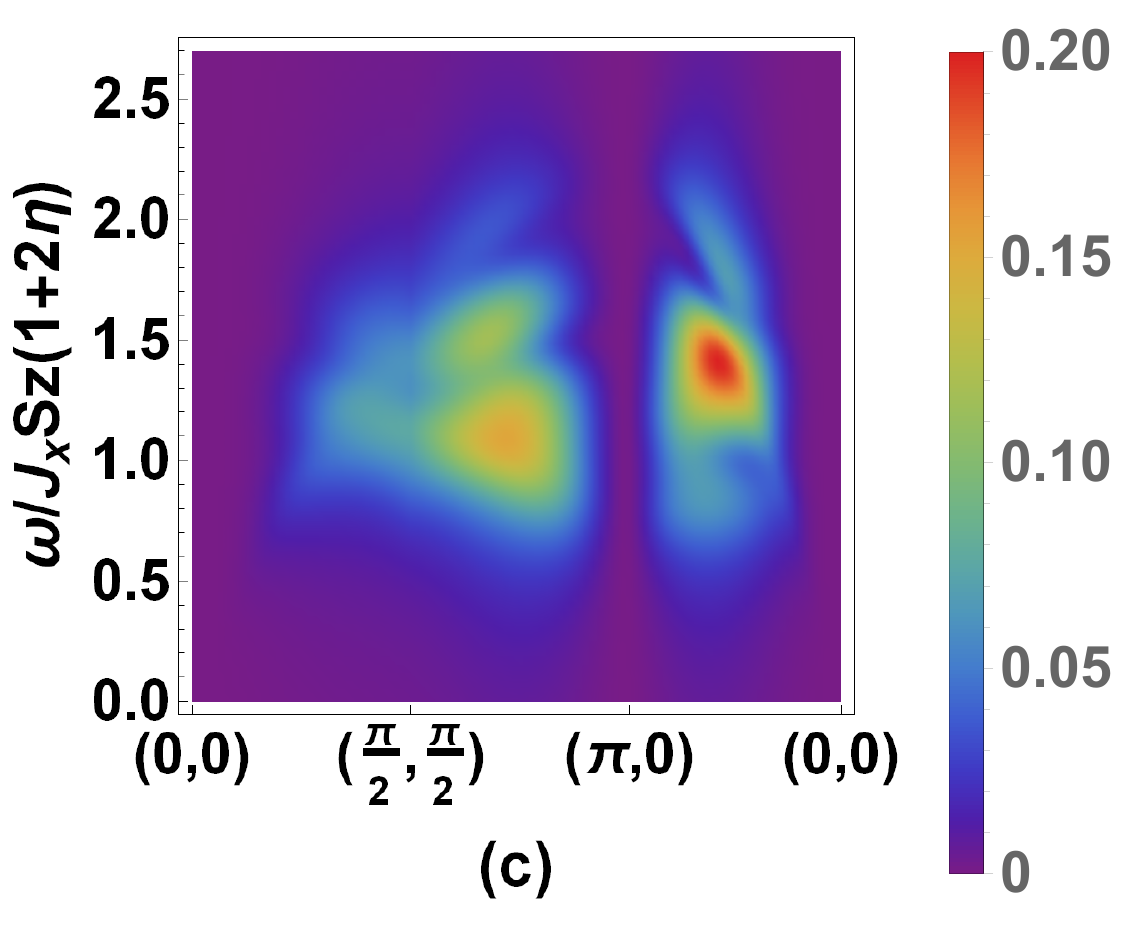}}\label{fig:8c}}
\caption{(Color online) RIXS intensity in CAF phase, with $\zeta=0.9, \eta=1, g_{x}=0.28, \Lambda_{\zeta}=0.9, \Lambda_{\eta}=0.7, \Omega^{ph}=0.15 E_{1}, \Delta_{k}=0.1 E_{1}$. (a) Damped bimagnon RIXS intensity, (b) magnon-phonon-magnon intensity, (c) total interacting intensity taking into account the effect of phonons.}
\label{fig:fig8}
\end{figure}

\begin{figure}[t]
\centering
\includegraphics[scale=0.2]{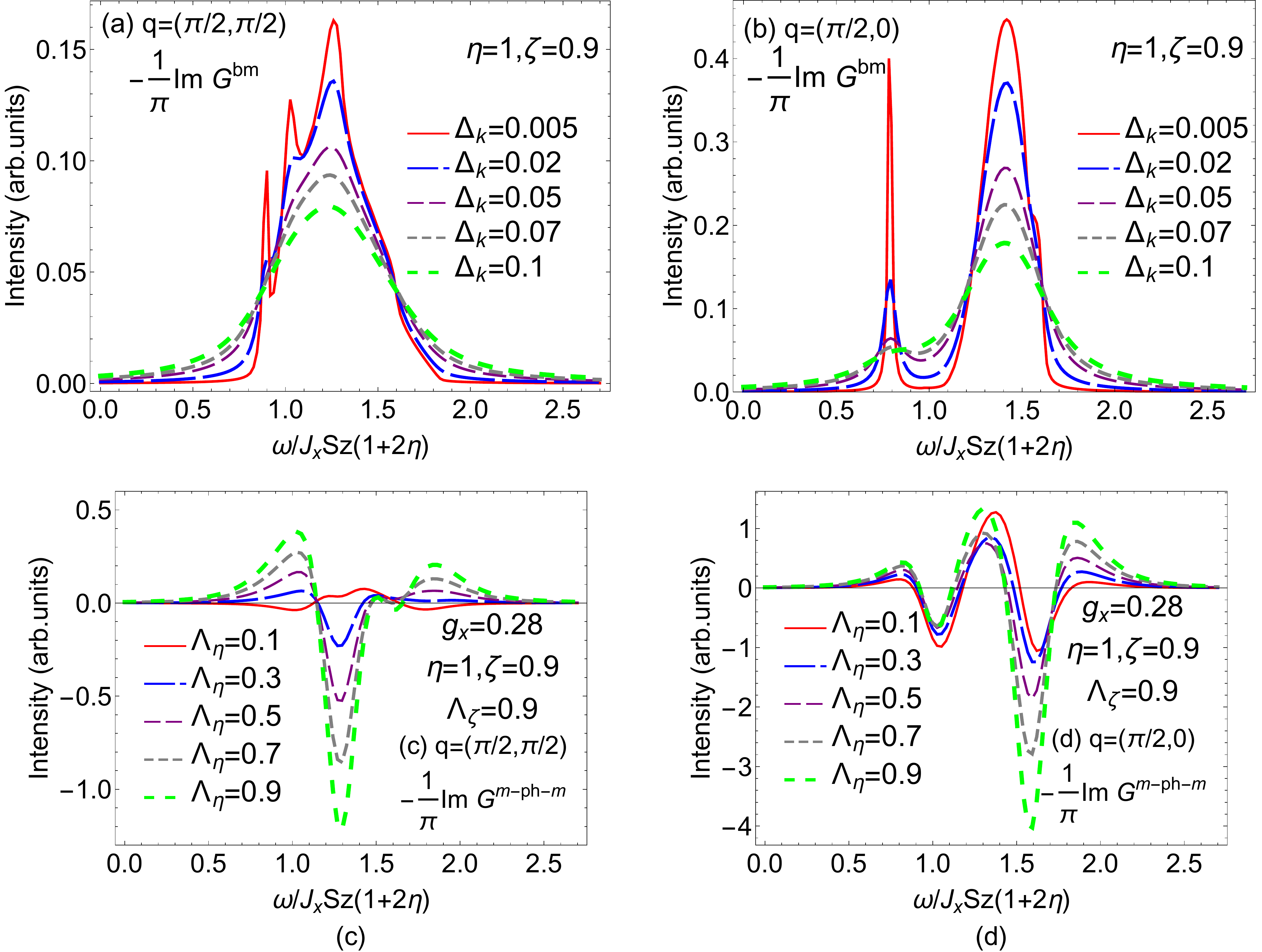}
\label{fig:9}
\caption{(Color online) CAF phase damping effects at the K$:\left(\frac{\pi}{2}, \frac{\pi}{2}\right)$ and the Y:$\left(\pi /2, 0 \right)$ points. $\Delta_{k}=0.1E_{1}$, $\Omega_{ph}=0.15E_{1}$, with $E_{1}=J_{x}Sz(1+2\eta)$ in the CAF phase. (a) -- (b) interacting bimagnon contribution. (c) -- (d) magon--phonon--magnon contribution.}
\label{fig:fig9}
\end{figure}

\begin{figure}[t]
\centering
\includegraphics[scale=0.2]{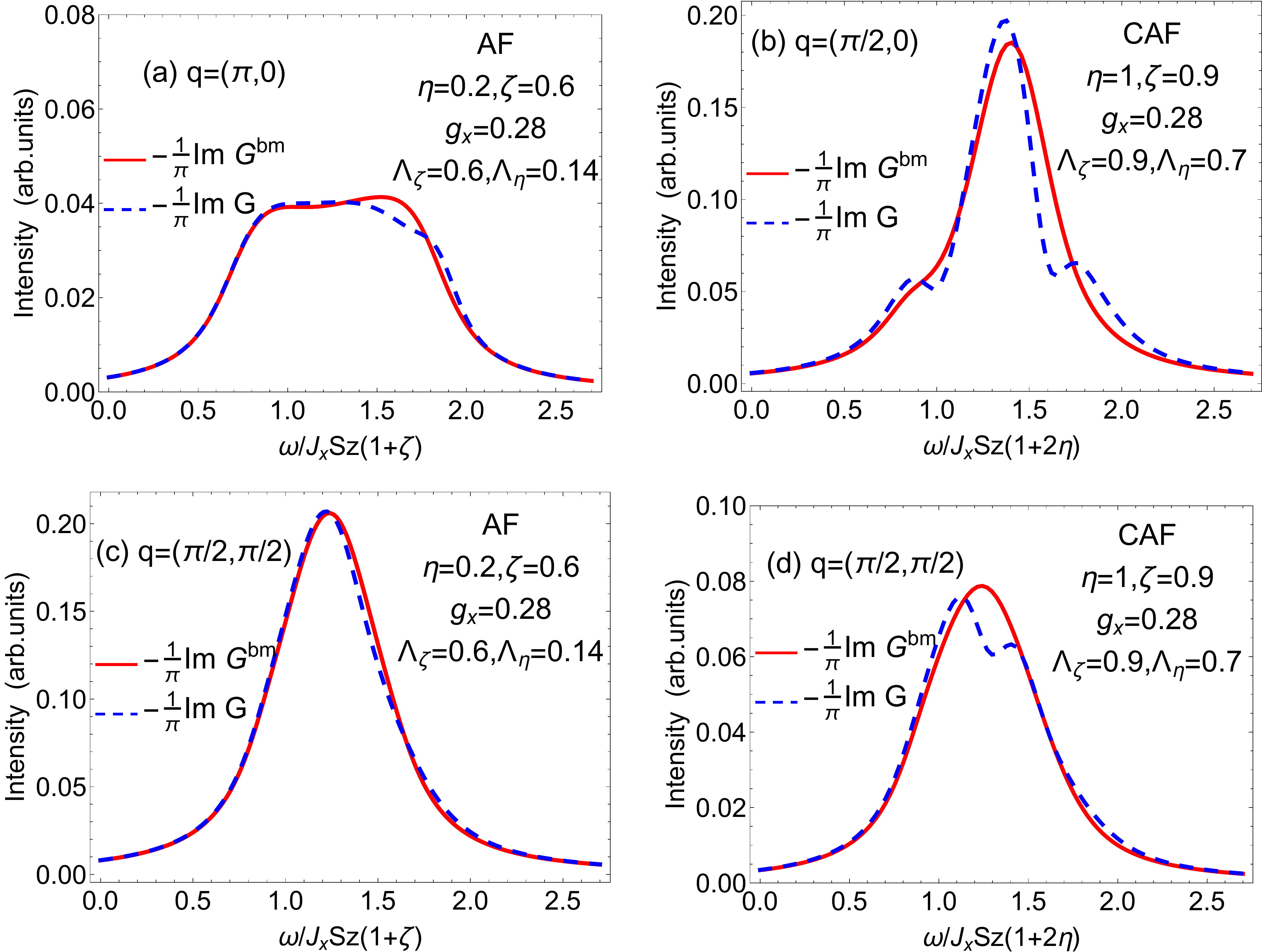}
\caption{(Color online) (a),~(c) AF phase interacting RIXS intensity with anisotropy effects. (b),~(d) CAF phase interacting RIXS line plots with anistropy effects. $\Delta_{k}=0.1 E_{1}$ and $\Omega_{ph}=0.15 E_{1}$, where $E_{1}$ takes the expression appropriate either the AF or the CAF phase.}
\label{fig:fig10}
\end{figure}

The 2D quantum Heisenberg AF can support a collinear AF phase for relevant magnetic interaction parameters ~\cite{PhysRevB.82.144407}. While the AF phase provides information on the parent magnetic compunds of cuprates, the CAF phase magnets are relevant in understanding how pnictide superonductivity can arise. To provide a comprehensive understanding of the model under study, in Fig.~\ref{fig:fig8}, we display our findings for the CAF phase. Compared to Fig.~\ref{fig:fig7} for the AF phase where the RIXS intensity peaks at the  K$:\left(\frac{\pi}{2}, \frac{\pi}{2}\right)$ point, in the CAF phase the RIXS spectral intensity weight attains its maximum value near $\left(\frac{\pi}{2}, 0\right)$. Futher comparison of the magnon-phonon-magnon RIXS plots in the two phases, Fig.~\ref{fig:fig8}(b) and Figs.~\ref{fig:fig5}(e) -~\ref{fig:fig5}(h), suggest that the magnon-phonon interaction creates additional satellite structures in the CAF phase for suitable interaction parameters. The spectral intensity of the magnon-phonon crorrelation is shown in  Fig.~\ref{fig:fig8}(b) where we notice that the intensity is spread over a much wider region in the BZ compared to the AF phase. The additional ripples in the CAF phase spectrum arises from the magnon-phonon-magnon spectrum. Systematic studies of the damping effect are displayed in Fig.~\ref{fig:fig9}. Similar to the AF phase, with increasing damping the RIXS bimagnon curves are broadened out. For the magnon-phonon-magnon plots, strong damping tends to suppress the satellite structures in the spectrum more at the K point, compared to the one at the Y point.
{\subsection{Anisotropy, frustration, and magnon-phonon coupling}\label{Subsec:anisfrusmagph}}
In Fig.~\ref{fig:fig10}, we compare and contrast the effects of spatial anisotropy between the AF and the CAF phase to reveal subtle differences between the RIXS response. A prominent unique two-shoulder peak is seen to develop in the CAF phase, which for the AF phase has typically been absent for the physical parameter region that was investigated. While in the AF phase the effects of anisotropy overpower the magnon-phonon effects, in the CAF phase the phonon peak structures are evident. These differences are crucial in distinguishing between the two different types of magnetic ordering. We also point out that at the Y-point the CAF phase spectrum, the magnon-phonon coupling introduces peaks both at the low and high energy spectrum. This feature is a consequence of the multi-satellite spectral feature of the magnon-phonon spectrum as seen in Fig.~\ref{fig:fig9}.

%%%%%%%%%%%%%%%%%%%%%%%%%%%%%%%%%%%%%%%%%%%%%%%%%%%%%%%%%%%%%%%%
\section{Conclusion}\label{Sec:Conclu}
Although the theory of magnon-magnon interactions on RIXS spectrum is well established there has been no thorough investigation on the role of phonons on the bimagnon magnetic spectrum of RIXS in the insulating 2D quantum Heisenberg magnet. In particular, microscopic investigations of the interaction between spin and lattice degrees of freedom and its effects on the bimagnon excitation is completely missing. In this paper, we provide a comprehensive theory of magnon-phonon interaction effects in both the AF and CAF phase at the K-edge. We compute the RIXS spectrum including the effects of damping, magnon-phonon coupling both at the nn and nnn level in the AF and the CAF phase. A detailed comparison between the two spectra reveal stark differences in the location of the maximum intensity points, the nature of the effect of phonon on the two-peak structure, and the distribution of the indirect RIXS spectral weight. From the perspective of experiments, the local sensitivity of magnon-phonon correlation on the bimagnon spin dynamics should be clearly evident in the nature of the varying spectral shapes. The unique differences manifest when anisotropy and frustration is taken into account. The final outcome of the magnon-phonon effects in the indirect K-edge RIXS spectrum, in both the AF and the CAF phase, is an experimentally observable feature. Based upon our analysis we infer that the spectrum is a delicate balance between competing nearest and next-nearest neighbor magnon and magnon-phonon coupling strengths. Finally, we hope our work will inspire future experimental investigation on the hitherto unexplored consequences of phonon effects on bimagnon RIXS dynamics at the K-edge.
%%%%%%%%%%%%%%%%%%%%%%%%%%%%%%%%%%%%%%%%%%%%%%%%%%%%%%%%%%%%%%%%
\begin{acknowledgments}
We thank Mark Dean for useful discussions. T.D. acknowledges invitation, hospitality, and kind support from Sun Yat-Sen University. T. D. acknowledges funding support from Augusta University Scholarly Activity Award and from Sun Yat-Sen University Grant No. OEMT--2017--KF--06. Z. X. and D. X. Y. are support by NKRDPC-2017YFA0206203, NSFC-11574404, NSFC-11275279, NSFG-2015A030313176, Special Program for Applied Research on Super Computation of the NSFC-Guangdong Joint Fund (the second phase), Leading Talent Program of Guangdong Special Projects.
\end{acknowledgments}
\appendix
\section{Dyson Maleev, Fourier and Bogoliubov transformation}\label{app:dmftbv}
The standard two sublattice Dyson-Maleev representation used to bosonize the spin operator is given by
\begin{eqnarray}
 S_{i}^{+}&=&(2S)^{1/2}(1-a_{i}^{\dagger}a_{i}/2S)a_{i},\non\\S_{i}^{-}&=&(2S)^{1/2}a_{i}^{\dagger} \label{DMA},\\S_{i}^{z}&=&S-a_{i}^{\dagger}a_{i},\non
\end{eqnarray}
for the sublattice A, and
\begin{eqnarray}
 S_{j}^{+}&=&(2S)^{1/2}b_{j}^{\dagger}(1-b_{j}^{\dagger}b_{j}/2S),\non\\S_{j}^{-}&=&(2S)^{1/2}b_{j} \label{DMB},\\S_{j}^{z}&=&-S+b_{j}^{\dagger}b_{j},\non
\end{eqnarray}
for the sublattice B. In the above $a (a^{\dagger})$ represents the bosonic annihilation (creation) operator on sublattice A and $b(b^{\dagger})$ represents the bosonic annihilation (creation) operator on sublattice B. Introducing the Fourier transform of bosonic operator we have
\begin{eqnarray}
a_{i}&=&\sqrt{\frac{2}{N}} \sum_{\bold{k}} e^{i\bold{k} \cdot \bold{r}_{i}}a_{\bold{k}},\non\\
b_{j}&=&\sqrt{\frac{2}{N}} \sum_{\bold{k}} e^{i\bold{k} \cdot \bold{r}_{j}}b_{\bold{k}},
\end{eqnarray}
where N is the total number of sites. Using the Bogoliubov transformations
\begin{eqnarray}
a^{\dagger}_{\bold{k}}&=&u_{\bold{k}}\alpha^{\dagger}_{\bold{k}}+v_{\bold{k}}\beta_{-\bold{k}},\non\\ b_{-\bold{k}}&=&v_{\bold{k}}\alpha^{\dagger}_{\bold{k}} + u_{\bold{k}}\beta_{-\bold{k}}.
\end{eqnarray}
we diagonalize the quadratic terms in $H^{s}$. The expressions for $u_{\bold{k}}$ and $v_{\bold{k}}$ are given by
\begin{equation}
u_{\bold{k}}=\sqrt{\frac{1+\epsilon_{\bold{k}}}{2\epsilon_{\bold{k}}}},\quad v_{\bold{k}}=-sgn(\gamma(k))\sqrt{\frac{1-\epsilon_{\bold{k}}}{2\epsilon_{\bold{k}}}}=-u_{\bold{k}}x_{\bold{k}}.
\end{equation}
Further useful relations used in the calculation include $u_{k}^{2}+v_{k}^{2}=1/\epsilon_{k}$ ad $u_{k}v_{k}=-\gamma_{k}/(2\epsilon_{k})$. The additional $\Delta_{A_{1}}$ Oguchi correction term $A_{k}$ ($A^{'}_{k}$) in the AF phase is given by
\begin{equation}
\begin{aligned}
&\frac{2}{N}\sum_{2}\gamma_{1}(1-2)u_{1}v_{1}u_{2}v_{2}\\
=&\sum_{2}\gamma_{1}(1)\gamma_{1}(2)u_{2}v_{2}+\Delta_{A_{1}}(\cos k_{1x}-\cos k_{1y})\frac{\gamma(1)}{2\epsilon_{1}},
\end{aligned}
\end{equation}
and in the CAF phase by
\begin{equation}
\begin{aligned}
&\frac{2}{N}\sum_{2}\gamma'_{1}(1-2)u'_{1}v'_{1}u'_{2}v'_{2}\\
=&\sum_{2}\gamma'_{1}(1)\gamma'_{1}(2)u'_{2}v'_{2}+\Delta'_{A_{1}}\cos k_{1x}(1-\cos k_{1y})\frac{\gamma'(1)}{2\epsilon'_{1}}.
\end{aligned}
\end{equation}

\section{Magnon-phonon vertex}\label{app:magph}
The spin-phonon vertex $A_{\lambda},\,B_{\lambda}, C_{\lambda},\,D_{\lambda}$ are short for $A_{\lambda}(k_{1},k_{2},q),\,B_{\lambda}(k_{1},k_{2},q),C_{\lambda}(k_{1},k_{2},q),\,D_{\lambda}(k_{1},k_{2},q)$. Their analytical expressions are given by
\begin{equation}
\begin{aligned}
\left(
\begin{array}{c}
A_{\lambda}\\
B_{\lambda}\\
C_{\lambda}\\
D_{\lambda}
\end{array}
\right)
=&\left(
\begin{array}{cccc}
u_{1}u_{2} & v_{1}v_{2} & u_{1}v_{2} & v_{1}u_{2}\\
v_{1}v_{2} & u_{1}u_{2} & v_{1}u_{2} & u_{1}v_{2}\\
v_{1}u_{2} & u_{1}v_{2} & v_{1}v_{2} & u_{1}u_{2}\\
u_{1}v_{2} & v_{1}u_{2} & u_{1}u_{2} & v_{1}v_{2}
\end{array}
\right)
\left(
\begin{array}{c}
\Gamma_{A}\\
\Gamma_{B}\\
\Gamma_{C}\\
\Gamma_{D}
\end{array}
\right)
\\
&+\Gamma_{E}
\left(
\begin{array}{c}
u_{1}u_{2}+\Phi_{G}\,v_{1}v_{2}\\
v_{1}v_{2}+\Phi_{G}\,u_{1}u_{2}\\
v_{1}u_{2}+\Phi_{G}\,u_{1}v_{2}\\
u_{1}v_{2}+\Phi_{G}\,v_{1}u_{2}
\end{array}
\right).
\end{aligned}
\end{equation}
For the AF phase we define
\begin{equation}
\begin{aligned}
&\Gamma_{A}=i(\chi_{x}\,\Delta_{x}(q)+\chi_{y}\,\Lambda_{\zeta}\,\Delta_{y}(q)),\\
&\Gamma_{B}=i(\chi_{x}\,\Delta_{x}(k_{1}-k_{2})+\chi_{y}\,\Lambda_{\zeta}\,\Delta_{y}(k_{1}-k_{2})),\\
&\Gamma_{C}=i[\chi_{x}(\Delta_{x}(k_{2}+q)-\Delta_{x}(k_{2}))+\chi_{y}\,\Lambda_{\zeta}(\Delta_{y}(k_{2}+q)-\Delta_{y}(k_{2}))],\\
&\Gamma_{D}=i[\chi_{x}(\Delta_{x}(k_{1})-\Delta_{x}(k_{1}-q))+\chi_{y}\,\Lambda_{\zeta}(\Delta_{y}(k_{1})-\Delta_{y}(k_{1}-q))],\\
&\Gamma_{E}=i\chi_{2}\,\Lambda_{\eta}(\Delta_{2}(k_{1})-\Delta_{2}(k_{2})-\Delta_{2}(q))
\end{aligned}
\end{equation}
where $\Delta_{x}(k),\Delta_{y}(k),\Delta_{2}(k)$ are short for $\Delta_{x}(q,\lambda,k),\Delta_{y}(q,\lambda,k)$ and $\Delta_{2}(q,\lambda,k)$ respectively,
\begin{small}
\begin{equation}
\begin{aligned}
&i\Delta_{x}(q,\lambda,k)=-\sum_{\delta_{x}}\hat{e}(q,\lambda)\cdot \hat{\delta_{x}}e^{-ik\cdot \delta_{x}},\\
&i\Delta_{y}(q,\lambda,k)=-\sum_{\delta_{y}}\hat{e}(q,\lambda)\cdot \hat{\delta_{y}}e^{-ik\cdot \delta_{y}},\\
&i\Delta_{2}(q,\lambda,k)=-\sum_{\delta_{2}}\hat{e}(q,\lambda)\cdot \hat{\delta_{2}}e^{-ik\cdot \delta_{2}}.
\end{aligned}
\end{equation}
\end{small}
and
\begin{small}
\begin{equation}
\begin{aligned}
&\chi_{x}=1-\frac{2}{NS}\sum_{p}(v_{p}^{2}+\cos p_{x}u_{p}v_{p}),\\
&\chi_{y}=1-\frac{2}{NS}\sum_{p}(v_{p}^{2}+\cos p_{y}u_{p}v_{p}),\\
&\chi_{2}=1-\frac{2}{NS}\sum_{p}(1-\cos k_{px}\cos k_{py})v_{p}^{2},
\end{aligned}
\end{equation}
\end{small}
For the CAF phase, we define
\begin{equation}
\begin{aligned}
&\Gamma_{A}=i(\chi'_{x}\,\Delta_{x}(q)+\chi'_{2}\,\Lambda_{\eta}\,\Delta_{2}(q)),\\
&\Gamma_{B}=i(\chi'_{x}\,\Delta_{x}(k_{1}-k_{2})+\chi'_{2}\,\Lambda_{\eta}\,\Delta_{2}(k_{1}-k_{2})),\\
&\Gamma_{C}=i[\chi'_{x}(\Delta_{x}(k_{2}+q)-\Delta_{x}(k_{2}))+\chi'_{2}\,\Lambda_{\eta}(\Delta_{2}(k_{2}+q)-\Delta_{2}(k_{2}))],\\
&\Gamma_{D}=i[\chi'_{x}(\Delta_{x}(k_{1})-\Delta_{x}(k_{1}-q))+\chi'_{2}\,\Lambda_{\eta}(\Delta_{2}(k_{1})-\Delta_{2}(k_{1}-q))],\\
&\Gamma_{E}=i\chi'_{y}\,\Lambda_{\zeta}(\Delta_{y}(k_{1})-\Delta_{y}(k_{2})-\Delta_{y}(q))
\end{aligned}
\end{equation}
where $\Delta_{x}(k),\Delta_{y}(k),\Delta_{2}(k)$ are the same as in AF phase. But,
\begin{equation}
\begin{aligned}
&\chi^{\prime}_{x}=1-\frac{2}{NS}\sum_{p}(v_{p}^{\prime 2}+\cos p_{x}u^{\prime}_{p}v^{\prime}_{p}),\\
&\chi^{\prime}_{y}=1-\frac{2}{NS}\sum_{p}(1-\cos k_{py})v_{p}^{\prime 2},\\
&\chi^{\prime}_{2}=1-\frac{2}{NS}\sum_{p}(v_{p}^{\prime 2}+u^{\prime}_{p}v^{\prime}_{p}\cos k_{px}\cos k_{py}).
\end{aligned}
\end{equation}

\section{Vertices and $\Gamma$ matrix} \label{app:vertgamma}
The vertex $u_{1}u_{2}u_{3}u_{4}\,V^{(4)}_{1234}$ can be transformed into a separable form (see Eq. \ref{decom}) with 18 channels and a q-dependent $\Gamma(q)$ matrix. The definition of these channels are given in Table \ref{tab:afvertex} \cite{LuoPhysRevB.92.035109}.
\begin{table}
\caption{Definition of the channels $v_\mathrm{n}(\kk)$ for AF phase and CAF phase}
\begin{ruledtabular}
\label{tab:afvertex}
\begin{tabular}{cll}
$n$ &  AF $v_{n}(\kk)$  & CAF $v'_{n}(\kk)$\\ \hline
 1 & $u_{\kk+\qq} u_{\kk} \cos k_x$         & $u'_{\kk+\qq} u'_{\kk} \cos k_x$\\
 2 & $u_{\kk+\qq} u_{\kk} \sin k_x$         & $u'_{\kk+\qq} u'_{\kk} \sin k_x$\\
 3 & $u_{\kk+\qq} u_{\kk} \cos k_y$         & $u'_{\kk+\qq} u'_{\kk} \cos k_x\cos k_y$\\
 4 & $u_{\kk+\qq} u_{\kk} \sin k_y$         & $u'_{\kk+\qq} u'_{\kk} \sin k_x\cos k_y$\\
 5 & $u_{\kk+\qq} v_{\kk}$                  & $u'_{\kk+\qq} u'_{\kk} \cos k_x\sin k_y$ \\
 6 & $v_{\kk+\qq} u_{\kk}$                  & $u'_{\kk+\qq} u'_{\kk} \sin k_x\sin k_y$\\
 7 & $v_{\kk+\qq} v_{\kk} \cos k_x$         & $u'_{\kk+\qq} v'_{\kk}$ \\
 8 & $v_{\kk+\qq} v_{\kk} \sin k_x$         & $v'_{\kk+\qq} u'_{\kk}$\\
 9 & $v_{\kk+\qq} v_{\kk} \cos k_y$         & $v'_{\kk+\qq} v'_{\kk} \cos k_x$\\
10 & $v_{\kk+\qq} v_{\kk} \sin k_y$         & $v'_{\kk+\qq} v'_{\kk} \sin k_x$\\
11 & $u_{\kk+\qq} v_{\kk} \cos k_x\cos k_y$ & $v'_{\kk+\qq} v'_{\kk} \cos k_x\cos k_y$\\
12 & $u_{\kk+\qq} v_{\kk} \sin k_x\cos k_y$ & $v'_{\kk+\qq} v'_{\kk} \sin k_x\cos k_y$\\
13 & $u_{\kk+\qq} v_{\kk} \cos k_x\sin k_y$ & $v'_{\kk+\qq} v'_{\kk} \cos k_x\sin k_y$\\
14 & $u_{\kk+\qq} v_{\kk} \sin k_x\sin k_y$ & $v'_{\kk+\qq} v'_{\kk} \sin k_x\sin k_y$\\
15 & $v_{\kk+\qq} u_{\kk} \cos k_x\cos k_y$ & $u'_{\kk+\qq} v'_{\kk} \cos k_y$ \\
16 & $v_{\kk+\qq} u_{\kk} \sin k_x\cos k_y$ & $u'_{\kk+\qq} v'_{\kk} \sin k_y$\\
17 & $v_{\kk+\qq} u_{\kk} \cos k_x\sin k_y$ & $v'_{\kk+\qq} u'_{\kk} \cos k_y$\\
18 & $v_{\kk+\qq} u_{\kk} \sin k_x\sin k_y$ & $v'_{\kk+\qq} u'_{\kk} \sin k_y$\\
\end{tabular}
\end{ruledtabular}
\end{table}
The non-zero matrix elements of $\Gamma(q)$ in AF phase are given by
\begin{equation*}
\begin{aligned}
&\Gamma(q)_{1,1}=-\theta,\quad \Gamma(q)_{1,5}=-\theta,\quad \Gamma(q)_{2,2}=-\theta,\quad \Gamma(q)_{3,3}=-\phi,\\
&\Gamma(q)_{3,5}=-\phi,\quad \Gamma(q)_{4,4}=-\phi,\quad \Gamma(q)_{5,5}=2\eta\,\theta\,\gamma_{2}(q),\\
&\Gamma(q)_{5,6}=-\gamma_{1}(q)/S,\quad \Gamma(q)_{5,7}=-\theta\,\cos q_{x}, \\
&\Gamma(q)_{5,8}=\theta\,\sin q_{x},\quad \Gamma(q)_{5,9}=-\phi\,\cos q_{y},\\
&\Gamma(q)_{5,10}=\phi\,\sin q_{y},\quad \Gamma(q)_{5,11}=-2\eta\,\theta\,\gamma_{2}(q),\\
&\Gamma(q)_{5,12}=2\eta\,\theta\,\gamma_{2}^{sc}(q),\quad \Gamma(q)_{5,13}=2\eta\,\theta\,\gamma_{2}^{cs}(q),\\ &\Gamma(q)_{5,14}=-2\eta\,\theta\,\gamma_{2}^{ss}(q),\quad \Gamma(q)_{6,1}=-\theta\,\cos q_{x},\\
&\Gamma(q)_{6,2}=\theta\,\sin q_{x},\quad \Gamma(q)_{6,3}=-\phi\,\cos q_{y},\quad \Gamma(q)_{6,4}=\phi\,\sin q_{y},\\
&\Gamma(q)_{6,5}=-\gamma_{1}(q)/S,\quad \Gamma(q)_{6,6}=2\eta\,\theta\,\gamma_{2}(q),\\
&\Gamma(q)_{6,15}=-2\eta\,\theta\,\gamma_{2}(q),\quad \Gamma(q)_{6,16}=2\eta\,\theta\,\gamma_{2}^{sc}(q),\\ &\Gamma(q)_{6,17}=2\eta\,\theta\,\gamma_{2}^{cs}(q),\quad \Gamma(q)_{6,18}=-2\eta\,\theta\,\gamma_{2}^{ss}(q),\\
&\Gamma(q)_{7,6}=-\theta,\quad \Gamma(q)_{7,7}=-\theta,\quad \Gamma(q)_{8,8}=-\theta,\quad \Gamma(q)_{9,6}=-\phi,\\
&\Gamma(q)_{9,9}=-\phi,\quad \Gamma(q)_{10,10}=-\phi,\quad \Gamma(q)_{11,5}=-2\eta\,\theta,\\
&\Gamma(q)_{11,11}=2\eta\,\theta,\quad \Gamma(q)_{12,12}=2\eta\,\theta,\quad \Gamma(q)_{13,13}=2\eta\,\theta,\\
&\Gamma(q)_{14,14}=2\eta\,\theta,\quad \Gamma(q)_{15,6}=-2\eta\,\theta,\quad \Gamma(q)_{15,15}=2\eta\,\theta,\\
&\Gamma(q)_{16,16}=2\eta\,\theta,\quad \Gamma(q)_{17,17}=2\eta\,\theta,\quad \Gamma(q)_{18,18}=2\eta\,\theta.
\end{aligned}
\end{equation*}
where
\begin{equation}
\begin{aligned}
&\theta=\frac{1}{S(1+\zeta)},\quad \phi=\frac{\zeta}{S(1+\zeta)},\\
&\gamma_{2}^{sc}(q)=\sin q_{x}\cos q_{y},\quad \gamma_{2}^{cs}(q)=\cos q_{x} \sin q_{y},\\
&\gamma_{2}^{ss}(q)=\sin q_{x}\sin q_{y}.
\end{aligned}
\end{equation}
The non-zero matrix elements of $\Gamma'(q)$ in CAF phase are given by
\begin{equation*}
\begin{aligned}
&\Gamma'(q)_{1,1}=-\theta',\quad \Gamma'(q)_{1,7}=-\theta',\quad \Gamma'(q)_{2,2}=-\theta',\\
&\Gamma'(q)_{3,3}=-2\eta\,\theta',\quad \Gamma'(q)_{3,7}=-2\eta\,\theta',\quad \Gamma'(q)_{4,4}=-2\eta\,\theta',\\
&\Gamma'(q)_{5,5}=-2\eta\,\theta',\quad \Gamma'(q)_{6,6}=-2\eta\,\theta',\quad \Gamma'(q)_{7,7}=\phi'\,\cos q_{y},\\
&\Gamma'(q)_{7,8}=-\gamma'_{1}(q)/S,\quad \Gamma'(q)_{7,9}=-\theta'\,\cos q_{x},\\
&\Gamma'(q)_{7,10}=\theta'\,\sin q_{x},\quad \Gamma'(q)_{7,11}=-2\eta\,\theta'\,\gamma_{2}(q),\\
&\Gamma'(q)_{7,12}=2\eta\,\theta'\,\gamma_{2}^{sc}(q),\quad \Gamma'(q)_{7,13}=2\eta\,\theta'\,\gamma_{2}^{cs}(q),\\
&\Gamma'(q)_{7,14}=-2\eta\,\theta'\,\gamma_{2}^{ss}(q),\quad \Gamma'(q)_{7,15}=-\phi'\,\cos q_{y},\\
&\Gamma'(q)_{7,16}=\phi'\,\sin q_{y},\quad \Gamma'(q)_{8,1}=-\theta'\,\cos q_{x},\\
&\Gamma'(q)_{8,2}=\theta'\,\sin q_{x},\quad \Gamma'(q)_{8,3}=-2\eta\,\theta'\,\gamma_{2}(q),\\
&\Gamma'(q)_{8,4}=2\eta\,\theta'\,\gamma_{2}^{sc}(q),\quad \Gamma'(q)_{8,5}=2\eta\,\theta'\,\gamma_{2}^{cs}(q),\\
&\Gamma'(q)_{8,6}=-2\eta\,\theta'\,\gamma_{2}^{ss}(q),\quad \Gamma'(q)_{8,7}=-\gamma'_{1}(q)/S,\\
&\Gamma'(q)_{8,8}=\phi'\,\cos q_{y},\quad \Gamma'(q)_{8,17}=-\phi'\,\cos q_{y},\\
&\Gamma'(q)_{8,18}=\phi'\,\sin q_{y},\quad \Gamma'(q)_{9,8}=-\theta',\quad \Gamma'(q)_{9,9}=-\theta',\\
&\Gamma'(q)_{10,10}=-\theta',\quad \Gamma'(q)_{11,8}=-2\eta\,\theta',\quad \Gamma'(q)_{11,11}=-2\eta\,\theta',\\
&\Gamma'(q)_{12,12}=-2\eta\,\theta',\quad \Gamma'(q)_{13,13}=-2\eta\,\theta',\\
&\Gamma'(q)_{14,14}=-2\eta\,\theta',\quad \Gamma'(q)_{15,7}=-\phi',\quad \Gamma'(q)_{15,15}=\phi',\\
&\Gamma'(q)_{16,16}=\phi',\quad \Gamma'(q)_{17,8}=-\phi',\quad \Gamma'(q)_{17,17}=\phi',\\
&\Gamma'(q)_{18,18}=\phi'.
\end{aligned}
\end{equation*}
where
\begin{equation}
\begin{aligned}
&\theta'=\frac{1}{S(1+2\eta)},\quad \phi'=\frac{\zeta}{S(1+2\eta)}.
\end{aligned}
\end{equation}
\bibliographystyle{apsrev4-1}
\newpage
% refphononrixs includes rixs + phonon references for square lattice, including general references to RIXS
\bibliography{refphononrixs}
\end{document}